*Attilio Sacripanti**Attilio Sacripanti*

# How to enhance effectiveness of Direct Attack Judo throws

*"Dr. Kano's dream : Judo rotational application"*

## Abstract

In this paper it is performed an appraisal of the Olympic Sport "Judo" effectiveness in the optics of Biomechanics, that is the Dr Kano's dream the rotational application of judo.
Kano wasn't able to develop his dream due to his premature death, but the biomechanical analysis is able to broaden the narrow translational vision of judo that is transmitted us by Kano's disciples.
Really speaking some learned followers of Dr. Kano like Kiuzo Mifune in Japan and Koizumi in England already had some rotational ideas, but few people appreciated their words.
To broaden the classical view biomechanics will use a very special field f experimentation.
This field of experimentation is obviously the high level competition in which most of these rotational application can be found applied more or less instinctively by high level Athletes.
Considering the two biomechanical tools that are the physical basis of judo throws it is possible to obtain such results from the analysis of high competition application:

Lever Techniques are enhanced in their effectiveness in three ways :
1. The rotational movements, strictly connected to the Lever techniques mechanics achieving victory (Ippon) in competition, can be extended to the unbalance phase (Kuzushi)
2. The rotational movements can be applied in a totally new way putting away even the unbalance that is basic in the Lever techniques.
3. The Lever tool can be hybridized with the application of a Couple to lower the energy consumption and to overcome some strong defensive resistance.

Couple Techniques are enhanced in their effectiveness also in three ways :
1. The Couple tool that, in principle, doesn't need unbalance allowing uke's body rotate around his center of mass, it is enhanced utilizing the Uke's body smaller resistance directions (normally summarized in Diagonal attacks).
2. The vertical rotational movements in the transverse plan with the axis in the sagittal plane can be added to the Couple application with Transverse Rotation ,and axis in the frontal plane to overcome some defensive resistance, mainly in the trunk and leg group of Couple techniques ( like Uchi Mata or O Soto Gari) .
3. The rotational movements can enhance the throwing action changing the inner mechanics of Couple into Lever applying a Torque, with the direction change of one force or the time delay of his application.

This analysis, expanded with a short physical framework, clarifies hidden deep concepts and helps teacher and coaches in their professional work.

# How to enhance effectiveness of Direct Attack Judo throws

*"Dr. Kano dream : Judo rotational application"*

*Pictures by Tamás Zaonhi & Gabriela Sabanu (IJF Archive ) Paco Lozano (EJU Archive)*
*Courtesy of IJF and EJU Presidents*

*Attilio Sacripanti*†‡§***

*ENEA (National Agency for Environment Technological Innovation and Energy) Robotic Laboratory
†University of Rome II "Tor Vergata", Italy
‡FIJLKAM Italian Judo Wrestling and Karate Federation
§European Judo Union Knowledge Commission Commissioner
**European Judo Union Education Commission Scientific Consultant

# How to enhance effectiveness of Direct Attack Judo throws

## "Dr. Kano dream : Judo rotational application"

### 1 Introduction

The dream of Dr. Kano (嘉納 治五郎）, as well known in the judo world, was to develop the traditional Kodokan (講道館）structure, along Ueshiba's (植芝盛平）rotational way; he wasn't able to develop his research in a rotational field due to his dead, but he paved the way to permit a natural Judo evolution with the Itsusu No Kata (五の形）, even it unfinished! [1]
Kano's scientific method was influenced by Ueshiba's Aikido (合氣道）, in effect Judo acquired, developed and adapted the rotational unbalance concept. Kusushi, Tsukuri phase became mainly rotary (but not officially into the books), taking a correct practical analysis of the throwing movement during competition into consideration.
Indeed, his student Kyuzo Mifune (三船久蔵）(10°dan) not only used to assert : "if the rival push you needs to rotate your body; if he pulls, you needs to shift against him in diagonal direction" , but even affirms into his book "Canon of Judo" pag.29 "… Essence of judo is to keep the center of physical gravity. Fall of a substance is in fact due to inability of maintenance of balance, so the form in which equilibrium is easily lost can be said "instability" . For instance, in Tachi-waza (stand trick), the most important is how to keep balance of body and to let the opponent to lose it. However, the further analytical study would convince you where is the center of two bodies grappled together and where is the leading movement created. Circling movements having his center firmly kept is a circle in plane view and a ball ( sphere ) when cubically view. The most perfect figure of all substances is a ball, and in mental image, too, very well-roundedness is indicative of superiority…" [2]
Rotational unbalance is very important to single out the importance of Tai Sabaki (体捌きLitt:, body shifting; body control) which must be considered, under this approach, in a most general view. In effect the rotational approach shows the evolution of judo effectiveness, the rotation, either in attack or in defense, is the base of an effective advanced Judo. Tai Sabaki, (体捌き) includes the whole Tori body's movements which will produce a rotational Kuzushi-Tsukuri (くずし-作り) phase.
Mister Koizumi（小泉軍治 ) 7°dan, in fact, professed: "the action to throw should be a continuous curved line…".
 And in page 44 of his book "My study of Judo" he underlines "… Feinting in the main is to produce the opponent's reaction by pushing or pulling in the opposite direction to that of the opening in view. In changing action from pushing to pulling , or from pulling to pushing, the movement should rhythmic and circular in order to retain the contact gained , plus the control over the opponent…" [3].
 From a didactics point of view it could be possible to define an attack Tai Sabaki, or a defense Tai Sabaki, but these are fictitious classifications, more often, of the same physical trajectory.

In general the continuous rhythmic circular movements of "Koizumi", and the unbalance directions, which are tangents to the " Mifune" circle developed, are infinite, like the Kano's rectilinear happo no Kuzushi. *Fig.1*

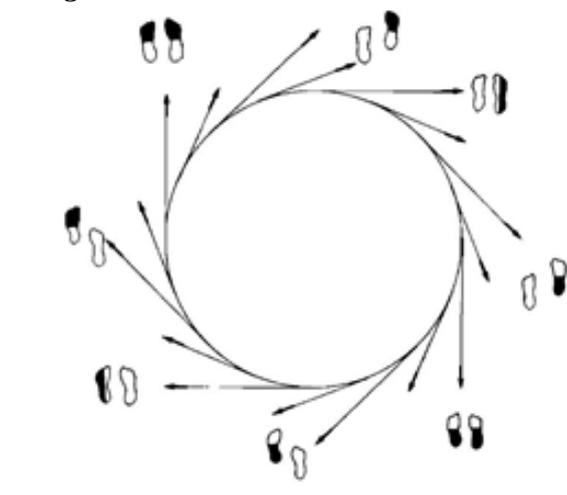

*Fig.1 Rotational Unbalances*

The introduction and execution of rotational Kuzushi-Tsukuri is the practical evolution and presupposes to be really applicable and effective, a condition of highly dynamic competition. Agonistic Judo, not very often, uses these rotational concepts as the most natural and appropriate to minimize the efforts and the energy, during the higher dynamic phases, in a modern competition. However this is not the outcome of a deep previous tactical study , but very often the result of fast movement in a quasi-chaotic situation.

Now it is important to insert two interesting notation about a wide part of the actual fighting style.
1. Normally during high level competition, for Tori (Attacker)the most utilized tactics are two: direct attack and action, reaction attack  ( for example during the World Championship 2010) more or less around the 75%.These two tactics are practically everywhere applied in front back direction or diagonal direction on a right line.
2. The second notation is about the shifting velocity of Couple of Athletes system, normally in high competitions the shifting velocity of Couple is very low and more often Tori applies the attack with very high speed, when Uke (Defender) is stationary or  proceeds slowly in right line (horizontal or vertical).

These notations inserted borne because the stationary situation, unconsciously preferred by the majority of athletes in attack, is due to the constant application during the technical training of the stationary Uchi Komi repetitions. From the other side the biomechanics assures us that during a fast dynamic phase a very little rotational moving, just a few degrees, permits to avoid often, the rival attack force, preventing each other throwing action (defensive Tai Sabaki); it's possible also, in such dynamic phase, to benefit by the insufficient rival kuzushi- tsukuri action, utilizing one's own unbalance to realize a combined effect between one's own kinetic energy and rival rotational deviation, in such way it is possible to apply a throws(attacking Tai Sabaki).

From the biomechanical point of view these subdivisions are useless cause they are too much metaphysical; the essence of the movement is, above all, the body's a rotation, using for example the fore-foot as pivot. In scientific words to study this rotational application, with a change of frame of reference, means  to complete the historical traditional  "translational approach" with the "rotational" one , that is related to the concept of better use of energy.

## 2 Aim of the work

The purpose of this paper is to study from the technical point of view the application of the rotational concept, energetically effective, to find, if it is possible, new technical variation or new chaotic techniques that can be applied in real competition. [4]

Normally athletes attackin three main way: **_Direct attack, Action-Reaction strategy_** and **_Combination of Throws_** this results show that the Action–reaction methods are the most useful , followed by Combination and Direct attack. In the next **_Diag. 1_** the results are shown:

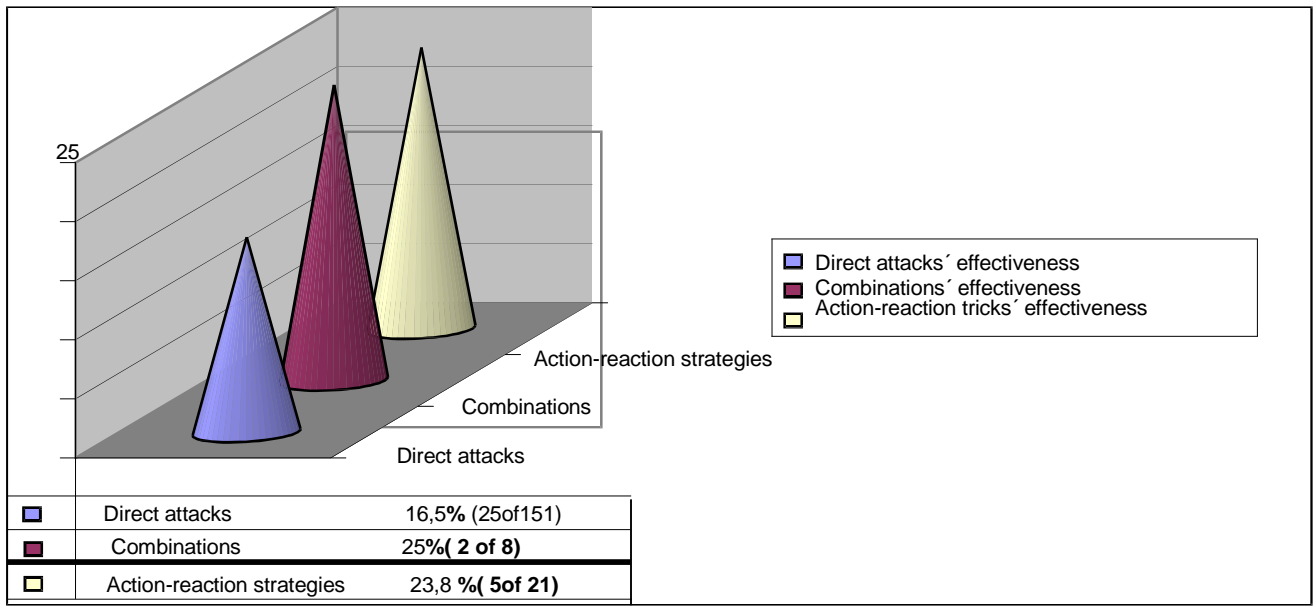

*Dig 1 Effectiveness of different types of attack* [5].

These three way to overcome the defensive capability of adversaries are the main practical ways applied in real competition.
In the world Championship 2010 happen in Japan the following table shows the same use of the same different attack types [6].**_Tab1_**

| Judo Attack Type in World championship 2010 Japan ||
|---|---|
| Attack Tactics | % |
| Direct Attack | 42.2 |
| Action Reaction | 34.8 |
| Counter Attack | 16.3 |
| Combination | 8,1 |

*Tab.1 Effectiveness of different types of attack WC Japan 2010* [6].

It is interesting to see the Japanese fighting Style utilized during the same world championship in the following table clearly appears that Japanese athletes with little change prefer the dynamic style couple techniques more than lever, with an interesting introduction more sutemi to overcome the fighting style of foreign people see **_Tab.2_**

| Judo Skill of Japanese Male team WC 2010 Japan ||
|---|---|
| Techniques | % |
| Ashi Waza | 33.3 |
| Te Waza | 13.3 |
| Koshi Waza | 6.6 |
| Sutemi Waza | 33.3 |
| Ne Waza | 12.6 |

*Tab.2 Judo Skill of Japanese Male teamWC2010 Japan* [6].

| Throws Effectiveness In London Olympic 2012 |||
|---|---|---|
| ***Throws*** | ***Effectiveness  Male %*** | ***Effectiveness  Female%*** |
| Seoi (Ippon– Morote-Eri) | 14.8      (329) | 8.2      (222) |
| UchiMata | 9.2      (138) | 15      (143) |
| O UchiGari | 15      (53) | 24      (49) |
| KoUchiGari | 12      (57) | 37      (35) |
| TaiOtoshi | 25      (36) | 23.8      (21) |
| Soto Makikomi | 10      (10) | 23.6      (17) |
| TaniOtoshi | 46      (13) | 50      (16) |
| UchiMata sukashi | 90      (10) | 100      (10) |
| ***Couple*** | ***28.7*** | ***39*** |
| ***Lever*** | ***24*** | ***26.4*** |

*Tab.3 Classical/Innovative  Throws Effectiveness In London  Olympic 2012*

Then in this work there are shown the way to improve the effectiveness of Judo throwing techniques on direct attack.

The introduction of new competitive techniques both of Couple or of Lever group, is connected with a more subtle way to fight with a different kind of goal to apply judo throws , with a sight both at energy saving and at biomechanical effectiveness determined by the actual difficulty for the body to resist a such attacks.

Enhance the technical effectiveness of judo throws in real competition is the whole result of biomechanical evaluation of  human body and right  way to apply forces in effective directions.

In fact the human body is unfit to hold out against such rotatory throwing  attacks, because muscular structure, body symmetry and joints' degrees of freedom are unable to defend him,  as well known, by diagonal or rotatory application of forces.

Biomechanical analysis of Judo basis and interaction in competition,  performed in several previous papers, can help effectively this effort for both groups of throwing techniques, because the rational foundation definite into them is useful to develop the right approach from the point of view both technical and scientific . [7,8,9,10]

## 3 Biomechanical overview of competition

In biomechanical term judo competition could be defined, as: an intriguing complex nonlinear system, with many chaotic and fractals aspects. Normally this complex system can be analyzed as every interacting physical system , dividing it in two main fields : Couple of athletes motion (system locomotion), and interaction between the two athletes ( throwing techniques).

### 3.1 Couple Motion

Into the Couple, human bodies of Athletes show complex responses connected to the human physiology whereof Brownian motion, starting from fractals till to multi-fractals aspects, is one of the basic modeling [11]. If we study, in deep, the change of position in time of bodies in space, starting from the motion of Centre of Mass in standing quiet position, till to Couple of Athletes ground track motion, Brownian Dynamics shows his ubiquitous presence in the motion description of this "Situation Sport"[12].
"Situation Sports" are identified as "sports in which the independence of simultaneous actions is not applicable for studying athlete's motion". Among these sports we can find both dual contact sports like judo and team one's. All they show the most complex motion during competition; in these specific field it is better analyzing motion in competition using a more sophisticated approach by means of statistical physics and chaos theory" [13]

### 3.2 Interaction into the Couple System

From the other side, interaction in competition may be analyzed by means of Classical Newtonian Physics, it's enhancing or changing is the main argument of this paper[14].
The mechanics of the Judo throws was widely analyzed in many previous papers [15,16,17]. Summarizing the structure it is based on two physical tools : the application on the adversary body of a Couple, a Lever or a their linear combination. The problem to shorten the distance between the athletes into the Couple of Athletes System ( for this Dual Sport) is solved always by three motion trajectories, the shorter and faster one's, called GAI ( General Action Invariant) plus a collision.
It is possible also to use different and longer trajectories, however GAI are the most energetically convenient.
After these common phases ( GAI+ Collision+ Tool [Couple or Lever]), Judo Interaction by means of throwing techniques splits in two main ways, depending by the physical tool utilized.
1. If the athlete applies as tool a Couple, after the right timing to apply the Couple, there is no other need, for example no unbalance, no stopping time and so on.
2. If the athlete applies as tool a Lever, he needs to finish off the technical action by a series of secondary musts, like stopping the adversary motion for a while, unbalancing him, positioning of the fulcrum and harms action ( application of forces in right direction).

For the second one, the must, is connected to the superior and inferior kinetic chains motion and right positioning connected to Kuzushi and Tsukuri phases.
Consequently, it is possible to resume the basic steps (useful for coaching and teaching) of jūdō interaction that occur during Dynamic Competitions.
These steps really reflect a continuous fluent movement:

1. *First: breaking the Symmetry to slow down the opponent (i.e., starting the unbalancing action)*
2. *Second: timing, i.e., applying the "General Action Invariant", with simultaneously overcoming the opponent's defensive grips resistance,*
3. *Third: sharp collision of bodies (i.e., the end of unbalance action)*
4. *Fourth:*
   *A. Application of "Couple of Forces" tool without any need of further unbalance action,*
   *B. Use of the appropriate "Specific Action Invariants", needing to increase unbalance action, stopping the adversary for a while, applying the "Lever" tool in Classic, Innovative or Chaotic way.*[18].

This steps represent the simplest movements which occur to throw the opponent.
Very often though, far more complex situations can arise under real fighting conditions. These complex situations which have evolved from the simple steps explained above depend on the combination of attack and defensive skills of both athletes.
However, the actual *Collision* step is very important for applying any real throwing technique
Then in function to both physical tools all judo throwing techniques can be grouped and classified as in the following ***tables.4,5***

| *Techniques Of Couple of forces* Couple applied by | Arms | Kuchiki daoshi    Uchi Mata Sukashi<br>Kibisu Gaeshi<br>Kakato Gaeshi<br>Te Guruma | *All Innovative Variation And Very few Chaotic form* |
|---|---|---|---|
| | Arm/s and leg | De Ashi Barai    O Uchi Gari<br>Okuri Ashi Barai    Ko Uchi Gake<br>Ko Uchi Barai    Ko Soto Gake<br>O Uchi Barai    Harai Tsurikomi Ashi<br>Tsubame Gaeshi    Yoko Gake<br>Ko Uchi gari    O Soto Gake<br>Ko Soto Gari    O Uchi Gake | |
| | Trunk and legs | O Soto Gari    O Tsubushi<br>O Soto gruruma    O Soto Otoshi<br>Uchi Mata    Ko Uchi Sutemi<br>Okurikomi Uchi Mata    Harai Makikomi<br>Harai Goshi    Ushiro Uchi Mata<br>Ushiro Hiza Ura Nage<br>Hane Goshi    Gyaku Uchi Mata<br>Hane Makikomi    Daki Ko Soto Gake<br>Yama Arashi    ( Khabarelli) | |
| | Trunk and arms | Morote Gari | |
| | Legs | Kani Basami | |

***Tab.4. Techniques based on a couple of forces*** [18].

| Techniques Of Physical lever Lever applied with | *Minimum Arm* (*fulcrum under Uke's waist*) | O Guruma          Ura Nage<br>Kata Guruma     Ganseki Otoshi<br>Tama Guruma     Uchi Makikomi<br>Obi Otoshi        Soto Makikomi<br>Tawara gaeshi    Momo Guruma<br>Makikomi    Kata sode Ashi Tsuri.<br>Sukui Nage        Daki Sutemi<br>Ushiro Goshi Utsuri Goshi | *All Innovative Variation and New (Chaotic) Forms* |
|---|---|---|---|
| | *Medium Arm* (*fulcrum under Uke's Knees*) | Hiza Guruma<br>Ashi Guruma<br>Hiza Soto Muso<br>Soto Kibisu Gaeshi | |
| | *Maximum Arm* (*fulcrum under Uke's malleola*) | Uki Otoshi        Yoko Guruma<br>Yoko Otoshi      Yoko Wakare<br>Sumi Otoshi       Seoi Otoshi<br>Suwari Otoshi     Hiza Seoi<br>Waki Otoshi      Obi Seoi<br>Tani Otoshi       Suso Seoi<br>Tai Otoshi        Suwari Seoi<br>Dai Sharin        Hiza Tai Otoshi<br>Ikkomi Gaeshi     Tomoe nage<br>Sumi Geshi        Ryo Ashi Tomoe<br>Yoko Kata Guruma   Yoko Tomoe<br>Uki Waza     Sasae Tsurikomi Ashi<br>Uke Nage     Morote Suwari Seoi | |
| | *Variable Arm* (*variable fulcrum from the waist To the knees*) | TsuriKomiGoshi        O Goshi<br>SasaeTsurikomi Goshi  Koshi Guruma<br>Ko Tsurikomi Goshi     Kubi Nage<br>O Tsurikomi Goshi      Seoi Nage<br>Sode tsurikomi Goshi   Eri Seoi Nage<br>Uki Goshi   Morote Seoi nage | |

*Tab.5. Techniques based on a physical lever* [18].

As it easy to see, techniques flowing from Couple tool are more or less half of the techniques flowing from Lever tool, this because biomechanically speaking the Couple throws are simplest than the Lever throws . Tori's body motions are both toneless and easier; some throws are applied by same body movement having the same biomechanical essence, even if Japanese called them in different way, treating them as different things. Japanese way to name these movements (as we see in the following pages remembering Kazuzo Kudo (工藤和三) ) is improper ; because, hiding the true physical basis, it brings about misunderstanding. Effectively if physical tools are understood by athletes, application, whatever grips or direction can arise, would be quite easy and natural.
In fact, if it is well understood how to apply Couple tool, then whatever is the hand position in grips the movement is only one, application of Couple into the three plane of symmetry; complex in these techniques is timing in application. Instead for Lever, whatever is forces direction by grips, one can apply stopping point in many positions, despite that, throwing mechanics is always the same. However in these techniques arms, legs and body coordination is complex.

*4 Classical , Innovative, and Chaotic throwing techniques*

In a previous paper, already was performed both the revision of Kano's Educational Kernel (Kuzushi, Tsukuri, Kake) , and the biomechanical analysis of judo throwing techniques , in this paper was for the first time studied the Judo Throws evolution in time around the world Tatami, defining these steps of changing from Classical Throws, to Innovative Throws, and to Chaotic Throws.

*Classical Throws : All the throwing movements as showed in the Kano's Go Kyo 1922, and Kodokan's Go Kyo 1985* [19]. *tab.6,7 Fig. 1,11*.

The last Kodokan Go Kyo or technical classification , Toshiro Daigo (醍醐敏郎）2005, is not considered because the Japanese author analyzes many variations of the Classical Techniques that in our definition are under the Innovative form of throws.

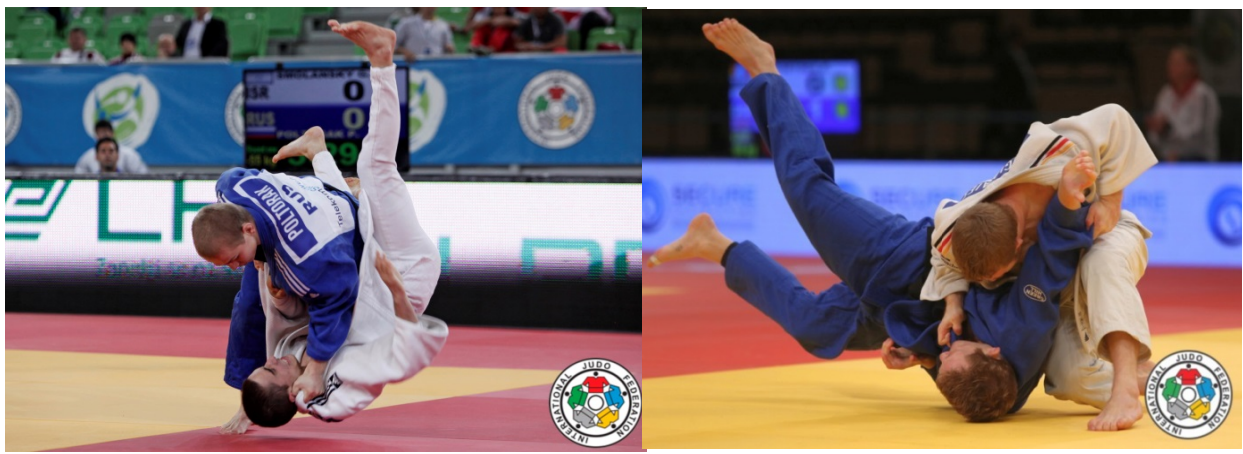

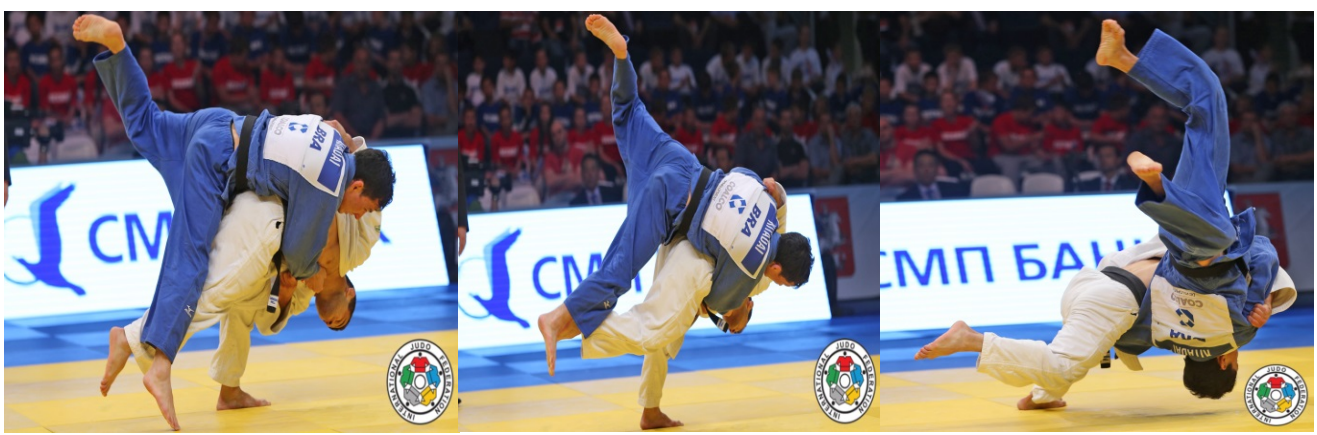

*Fig. 1,2, 3-5: Classical Throws : Lever Group- Morote Seoi Nage; Lever Group-Morote Seoi Otoshi ; Couple Group- a Rare example of competitive Hane Goshi .*

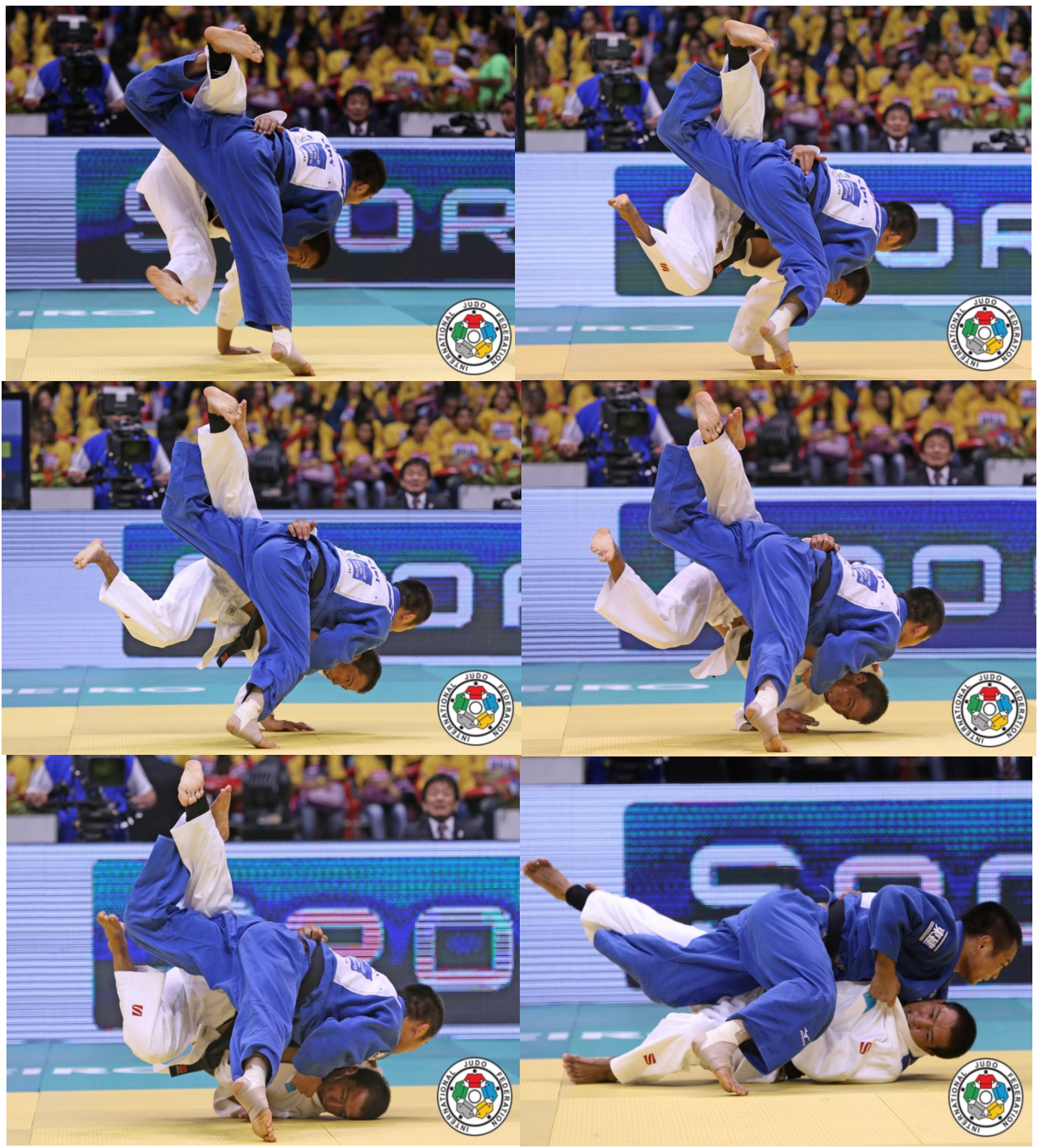

*Fig. 5-11: Couple Group- a perfect Classic Uchi Mata with the Tori's body motion only into the Sagittal plane.*

| Tachi Waza | Te Waza | Seoi nage<br>Tai otoshi<br>Kata guruma<br>Uki otoshi<br>Sumi otoshi<br>Sukui nage |
|---|---|---|
| | Koshi Waza | Uki goshi<br>O goshi<br>Koshi guruma<br>Utsuri goshi<br>Tsuri lomi goshi<br>Harai goshi<br>Tsuri goshi<br>Hane goshi<br>Ushiro goshi |
| | Ashi Waza | De ashi barai<br>Hiza guruma<br>Sasae tsurikomi ashi<br>O soto gari<br>O uchi gari<br>Ko soto gari<br>Ko uchi gari<br>Okuri ashi barai<br>Uchi mata<br>Ko soto gake<br>Harai tsurikomi ashi<br>Ashi guruma<br>O guruma |
| Sutemi Waza | Ma sutemi waza | Tomoe nage<br>Ura nage<br>Sumi gaeshi<br>Tawara gaeshi<br>Hikkikomi gaeshi |
| | Yoko sutemi waza | Uki waza<br>Yoko gake<br>Yoko guruma<br>Tani otoshi<br>Yoko wakare<br>Soto makikomi |

*Tab 6. Classical Techniques: the five group of Kodokan Technical Classification 1922* [19]

| Tachi Waza | Te Waza | Seoi nage<br>Tai otoshi<br>Kata guruma<br>Sukui nage<br>Uki otoshi<br>Sumi otoshi<br>Obi otoshi<br>Seoi otoshi<br>Yama arashi |
|---|---|---|
| | Ashi Waza | De ashi barai<br>Hiza guruma<br>Sasae tsurikomi ashi<br>O soto gari<br>O uchi gari<br>Ko soto gari<br>Ko uchi gari<br>Okuri ashi barai<br>Uchi mata<br>Ko soto gake<br>Ashi guruma<br>Harai tsurikomi ashi<br>O guruma<br>O soto guruma<br>O soto otoshi |
| | Koshi Waza | Uki goshi<br>O goshi<br>Koshi guruma<br>Tsuri lomi goshi<br>Harai goshi<br>Tsuri goshi<br>Hane goshi<br>Utsuri goshi<br>Ushiro goshi |
| Sutemi Waza | Ma sutemi waza | Tomoe nage<br>Ura nage<br>Sumi gaeshi<br>Hiki komi gaeshi<br>Tawara gaeshi |
| | Yoko sutemi waza | Yoko otoshi<br>Tani otoshi<br>Hane makikomi<br>Soto makikomi<br>Uki waza<br>Yoko wakare<br>Yoko gake<br>Daki wakare<br>Uchi makikomi |

*Tab 7. Classical Techniques: the five group of Kodokan Technical Classification 1985* [19]

To increase the effectiveness of throw, in real competition, athletes' bodies must collide each other. In effect because it is difficult if not impossible to fit in, during competition in the position that people train itself normally in *uchi komi* or *nage komi,* applying forces in classical directions, in that way borne the ***Innovative Techniques*** ( Roy Inmann), that are defined as [20,21,22,23]

*"Innovative Throws" are all throwing techniques that keep alive the formal aspect of Classic Jūdō throws, and differ in terms of grips and direction of applied forces only. Fig12,13*

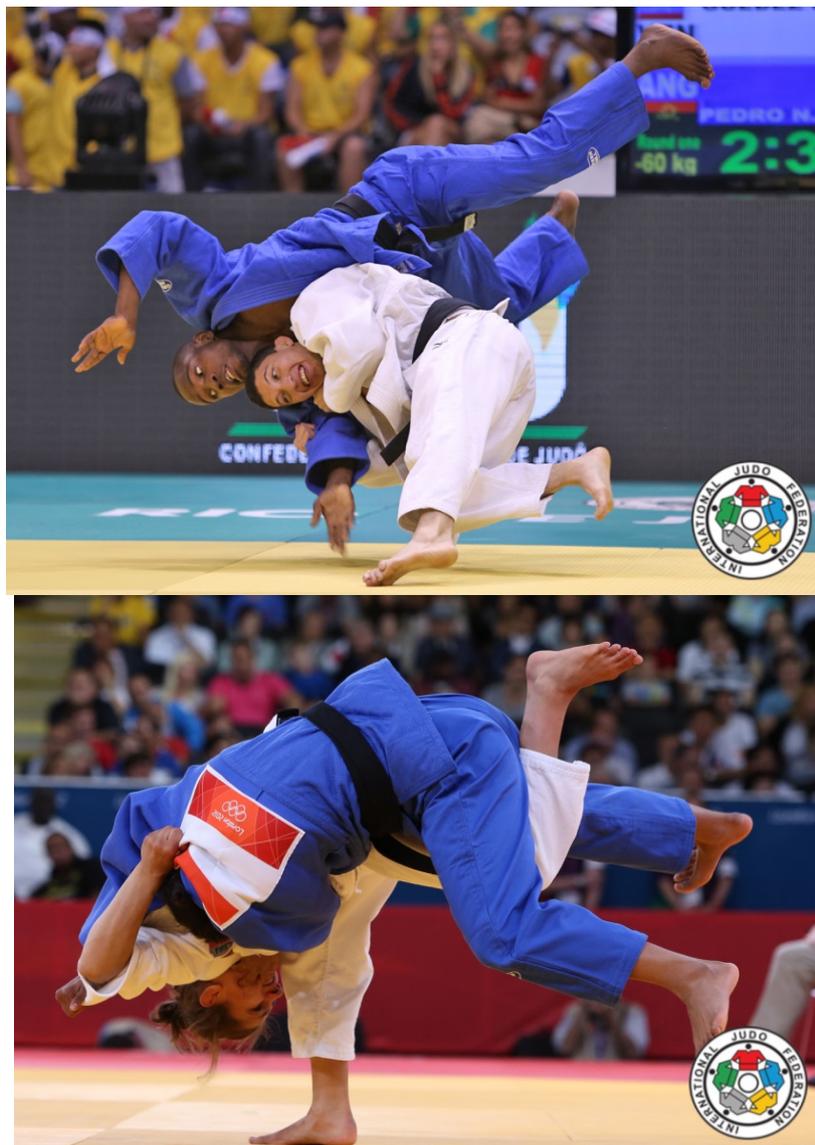

*Fig 12,13 Innovative Throws : Lever Group- Ippon Seoi Nge; Couple Group-Uchi Mata*

***Innovative Throws*** are variations (***henka)*** of classical *Kōdōkan* throwing techniques, which are either Couple of Force-type or Lever-type techniques biomechanically speaking, while it remains easy to still recognize a well-known basic traditional technique (40 or 47*gokyō* throwing techniques) in them.
However, there are other "non-classic" solutions applied in competition and which are different from 'Innovative'(*henka)* throws, which we define as: "***Chaotic Forms***".
Oftentimes these ***Chaotic Throws*** are mainly limited to class of lever group.
We non analyze these types of throws more in depth, then the real difference between the goals of kuzushi/tsukuri in both biomechanical groups of throws will be clarified.

***" Chaotic Throws" are characterized by the application of different grips positions which applying force in different (nontraditional) directions, while simultaneously applying (stopping points ) in non-classical position, or utilizing "no rational" shortening trajectories (longer than the usual) between athletes ( see Fig25 Reverse Seoi)***

These techniques are not named as the classical ones', and this is one of the greater difficulties to describe them in classical way. From there the question arises whether it is necessary to use new name for different throws ? About the names of classic jūdō throws it is appropriate to consider the words of Kazuzō Kudō (工藤和三) (9 Dan), one of the last students of Jigorō Kanō, in his book "Dynamic Judo Throwing technique" [24] he said :"*Jūdō names fall into the following categories:*
1. *Name that describe the action:* ***ō-soto-gari, de-ashi-barai, ō-uchi-gari-gaeshi.***
2. *Names that employ the name of the part of body used:* ***hiza-guruma, uchi-mata.***
3. *Names that indicate the direction in which you throw your opponent:* ***yoko-otoshi.***
4. *Names that describe the shape the action takes:* ***tomoe-nage*** ('tomoe' is a comma-shaped symbol).
5. *Names that describe feeling of the techniques:* ***yama-arashi, tani-otoshi.***

*Most frequently, jūdō techniques names will use the content of one or two of these categories.".*

Biomechanics instead gives not names but helps the understanding , clarifying the mechanical action of the tool utilized for these techniques, in such way, names are superfluous the main thing is to understand the inner mechanism and to apply it in whatever situation.
*Chaotic form of Throws: Fig 14-25*

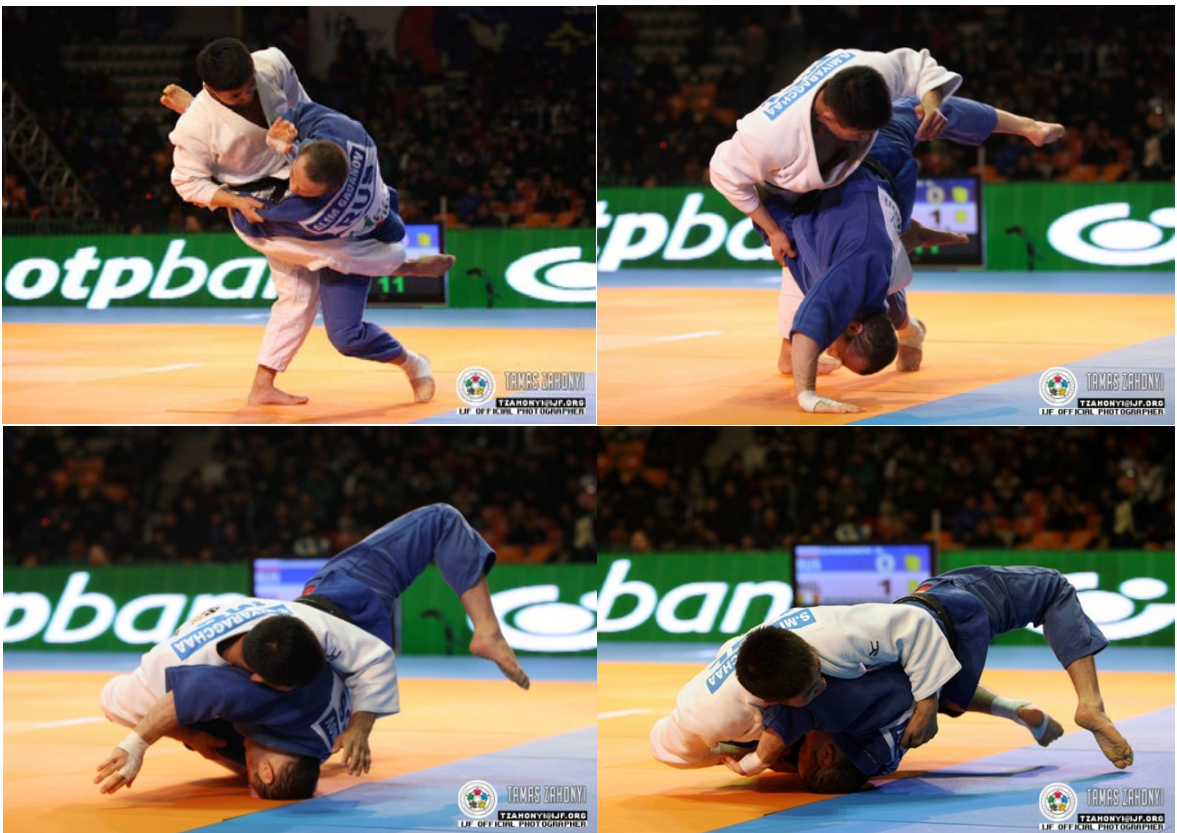

*Fig 14, 17  Chaotic Throws : Lever Group -No name;*

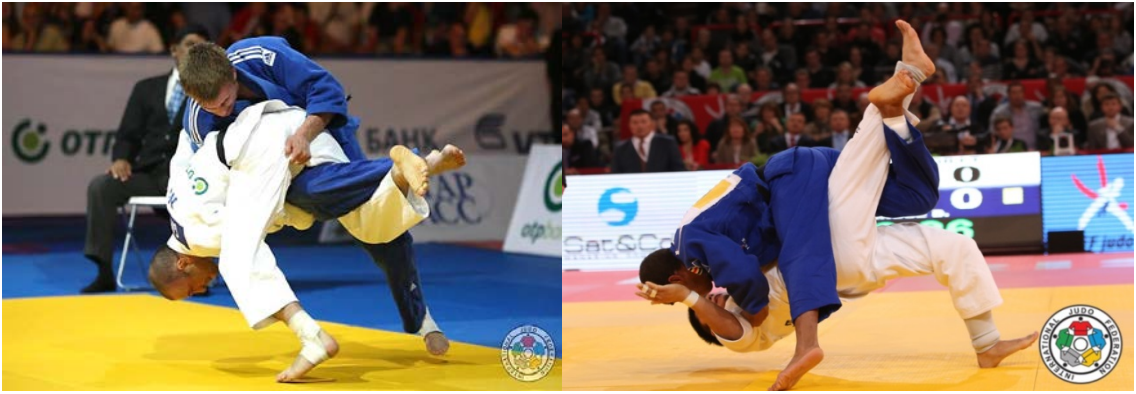
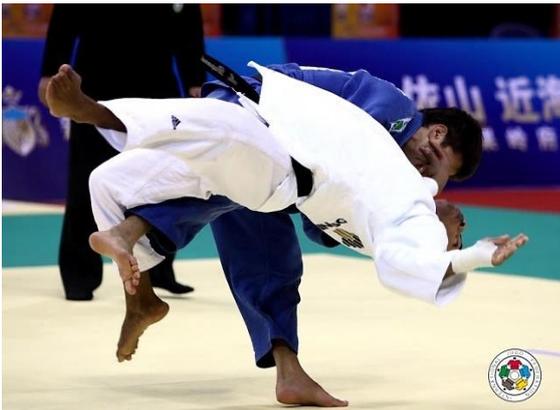
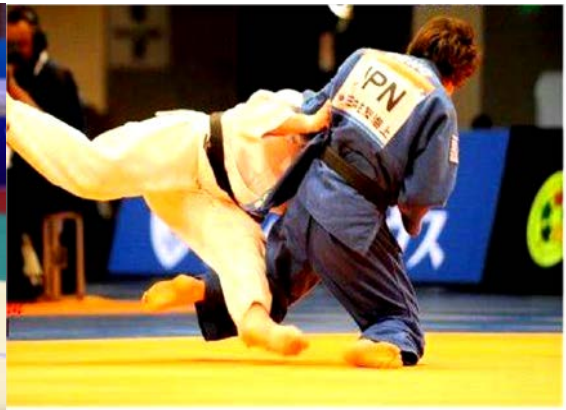
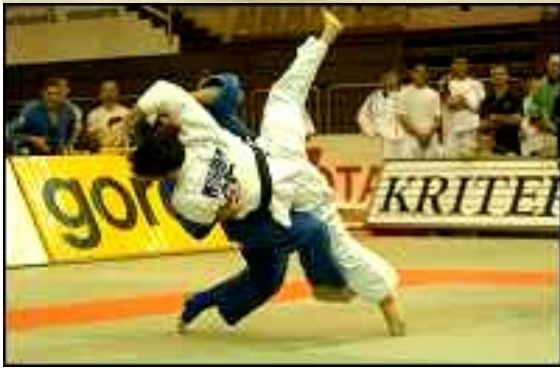
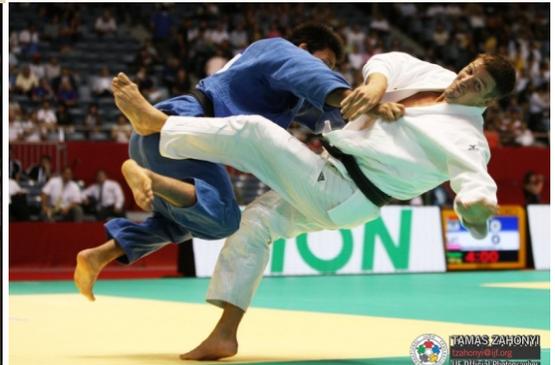
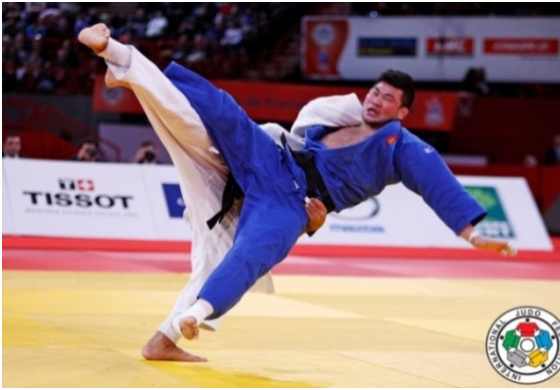
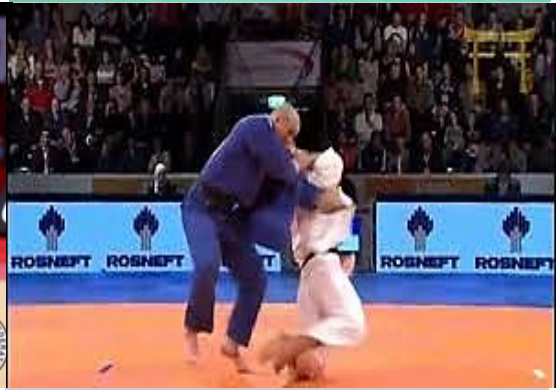

*Fig 18,19 Couple Group-No name; 20, 21 , 22 Lever Group-No name; 23, 24 Couple Group-No Name; 25Lever Group - Reverse Morote Seoi Nage.*

## 5 Rotational Enhancement

Angular movement is rotational movement. When an athlete rotates, it always turns around an axis. In judo fight the axis is always imaginary, like when the body rotates in the air, free of support or around him. In this last case, the axis of rotation passes through the exact center of gravity of the body.

The center of mass moves into the body, every time the body changes his shape.

If the arms are raised overhead, or one leg lifts up higher or trunk bends, Center of Mass moves away from the original position into the body. When the athlete's body is in the air rotating, it always rotates around its exact center no matter where that center is located.

The concepts of angular motion are similar to linear motion, but the terms change in order to specifically identify an association with rotation. It still takes a force to overcome inertia in order to produce momentum. The force that produces rotation is called Torque.

Torque is required in order to rotate a body, if on the athlete's body acts two forces parallel equal and opposites, we speak about Couple ( of forces) , also a Couple can rotate a body.

Instead of having to overcome simple inertia (weight or mass), in a rotational situation it must overcome angular inertia.

Not only the mass resists to the movement but also, when you are trying to turn a body, the length or the distance of the body has an effect on his resistance to turn.

The longer or most distant the body is, the more difficult it becomes to turn the body.

So in the rotational application of judo, there are two factors that constitute the angular inertia of a body–the mass and the length or distance of the body from the rotation axis.

In general there will be more angular inertia (more resistance to torque or angular force) the more mass the body contains and the longer it is when the torque or couple are applied.

Once a force (torque) of sufficient magnitude (enough to overcome the angular inertia) is applied to a body, angular momentum will be produced. The total amount of angular momentum will depend on the angular inertia (i.e., how much mass and how distant the body is) and the speed (angular velocity) the body is turning. It is important to understand how Torque is created.

In general terms athlete's body can rotate or by a combination of two forces (in physics terms – a Couple), or by applying what is called an eccentric force (i.e., a force that is not directed through the center of gravity of a body) using a lever system around a stopping point (fulcrum) .

It is important to realize that all the angular momentum variations are created while the adversary's body is still in contact with the ground, in other words, at the time of the takeoff. This angular momentum is a product of the angular inertia (mass and the distance of the body from the Tori body as active force) and the angular velocity created.

Once in the air, it is impossible for Uke to change his angular momentum.

The rotational application is both from the energetic and biomechanical point of view the most effective one's application of judo throws. It is well known, from judo biomechanics, that for the Couple Throwing techniques, that are energetically convenient respect to Lever group, ideally the Uke's body turns around his Center of Mass.

In effect if the athlete stands still the application of Couple on the body does not produce acceleration on the center of mass because:

$$a_{CM} = \frac{\sum F}{\sum m} \ but \ into \ a \ Couple \ is \ \sum F = 0 \ then \ a = \frac{0}{\sum m} = 0 \quad (1)$$

If the athlete moves himself, with constant velocity $v_{const}$, or acceleration β, neither velocity nor acceleration does not change with the application of a Couple on his body, but the body moves himself with the same velocity or acceleration rotating around his center of Mass.

On the other hand, during the Lever throws application both Center of Mass and athletes' body move in the space.
However rotation is essential and already present in classic judo throwing techniques, because the need to throw on the back the adversary for regulation, compels Tori to apply a final rotation or a complete rotational movement to obtain victory in competition. *Fig 26-29*

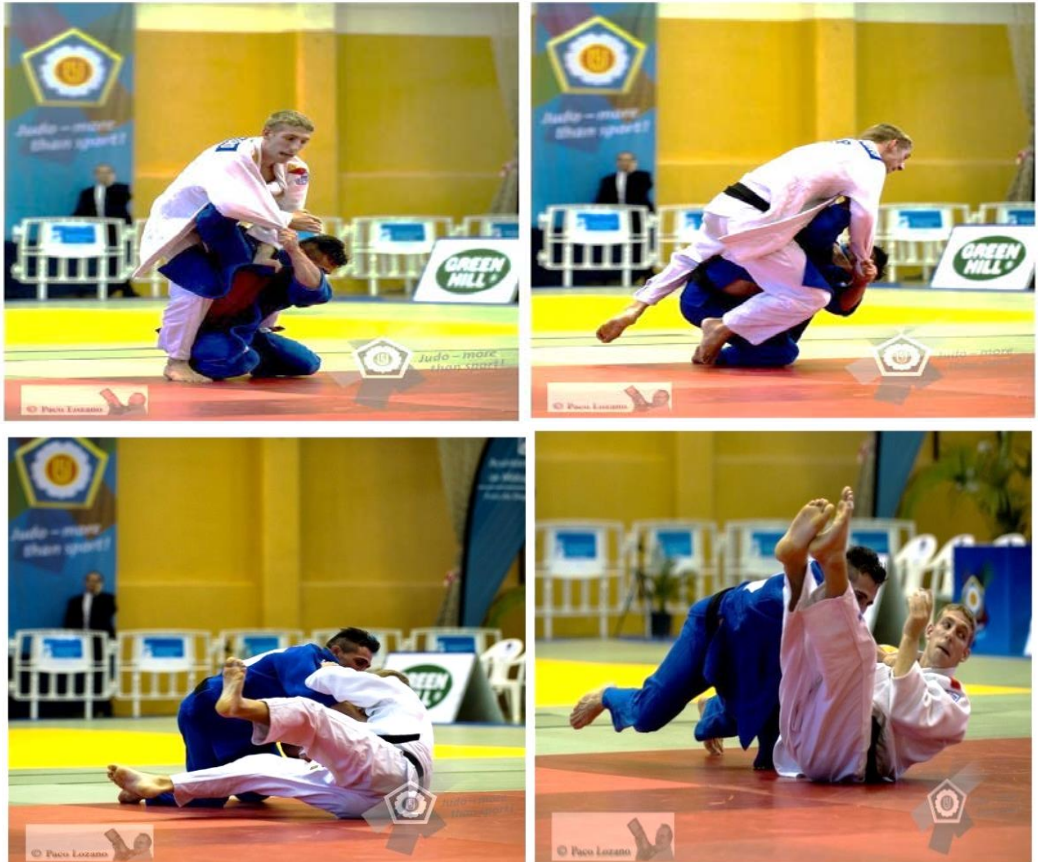

*Fig 26-29 Classical application of vertical rotatory movement for Ippon by Suwari Seoi Nage*

Rotational applications are useful also both to spend less energy and to overcome the natural defensive capabilities of the body.
The basis of this upset is grounded both on the mechanical conformation of the human body and the laws of Newtonian Physics.
It is interesting to highlight that the athlete's body is not able to defend him both from diagonal and rotational application of throwing forces.
The muscular structure is unable to resist to diagonal/rotational forces and human body is unable to resist so much, against such solicitations.
Some rotational application falls into the innovative and chaotic throw-building.
There too many way to apply rotational forces into the Couple of athletes system, perhaps this was the first stumbling block that stopped this kind of lengthen after Kano death.
The rotational complex could be organized on many bases, but the mechanical basis of judo throwing techniques helps us to choose the right way to build up these new Innovative or Chaotic techniques.
To apply useful rotational techniques it is essential to study the Center of Mass motion, then these new techniques must push the Center of Body Mass along circular or pseudo-circular paths.

## 6  Rotatory modification of Couple techniques

As previously stated rotatory movements are already present in the final part of some Couple throws, now it is interesting to analyze some rotatory variations that are utilized in competition to enhance the effectiveness of these techniques.
But what is the better way to single out the effective rotatory variation in Couple group?
- To consider the defensive inabilities of the adversaries' bodies!

The only right way, biomechanically speaking, is to evaluate the human body structure.
Analyzing human body structure means to select the direction in which the muscular structure is less able to resist to throwing force applied by Tori.
Single out these weakness helps to identify the trajectories of better use of the energy that can be converted in trajectories of the adversary's center of mass in space.
The last statement, considering the inner mechanics of Couple group techniques (rotation of Uke body around his center of mass) means to apply some further special rotation enhancing the effectiveness of throws.
One other way is to hybridize the couple mechanics into the lever by the non-simultaneous application of couple.
Resuming the rotational application of Couple Group techniques needs a severe change in their application and in their nature to enhance their effectiveness in some special situations.
During the application of throwing techniques it is important to remember the importance of Tori's head and neck in defining the throws direction, these movements are ruled by: vestibular system, proprioceptive neck reflexes, and mechanics of the system.
All throwing techniques of Couple groups increase their effectiveness in the rotatory meaning only with drastic changes in their application, or modifying direction, or applying special consecutive rotatory motion, or varying the inner mechanics of the techniques.
In the first, applying Couple in specific diagonal directions: *Fig 30-33*

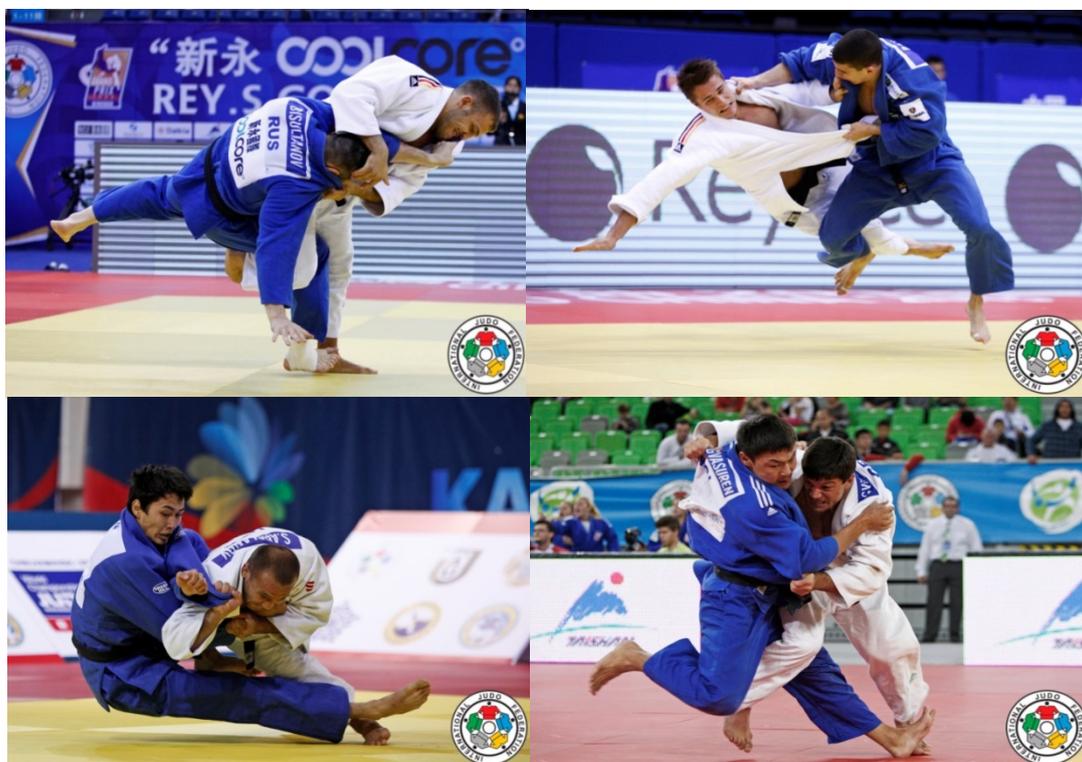

*Fig 30-33 Diagonal application of some Innovative Couple throws: two Osoto Gari, Ko Uchi gari, New Uchi Mata Henka (?).*

The second one is the application of consecutive rotation into the transverse (horizontal) symmetry plane to enhance the effectiveness specifically of Couple techniques applied by trunk and leg.
In this group it is possible to find two of the most utilized throws Uchi Mata and O Soto Gari; that biomechanically speaking are the same basic way to apply the Couple Tool, *Fig 34*; although Japanese Classic vision looks at two different movements or techniques.

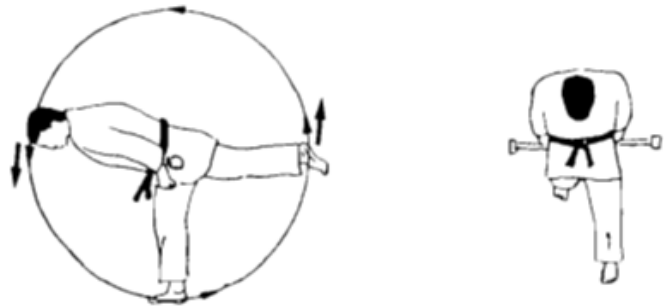

*Fig. 34 Basic and same way to apply the Couple in trunk leg group, like O Soto Gari and Uchi Mata the different names hail from Uke's front back difference*. [13]

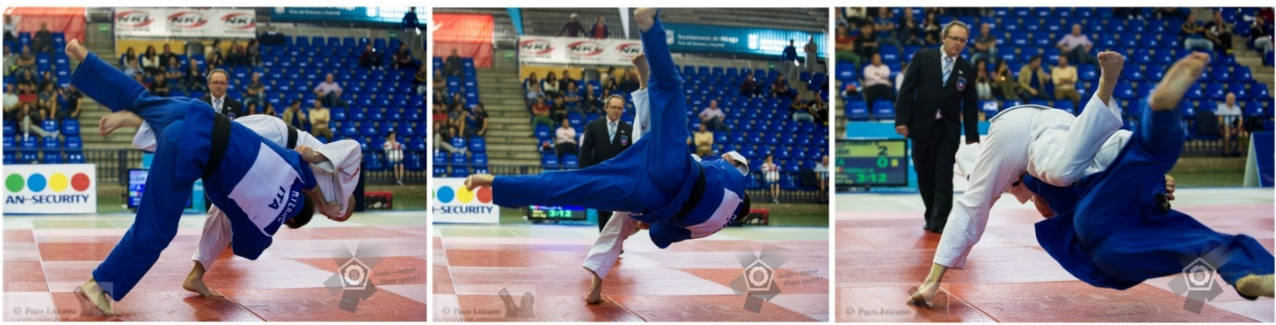

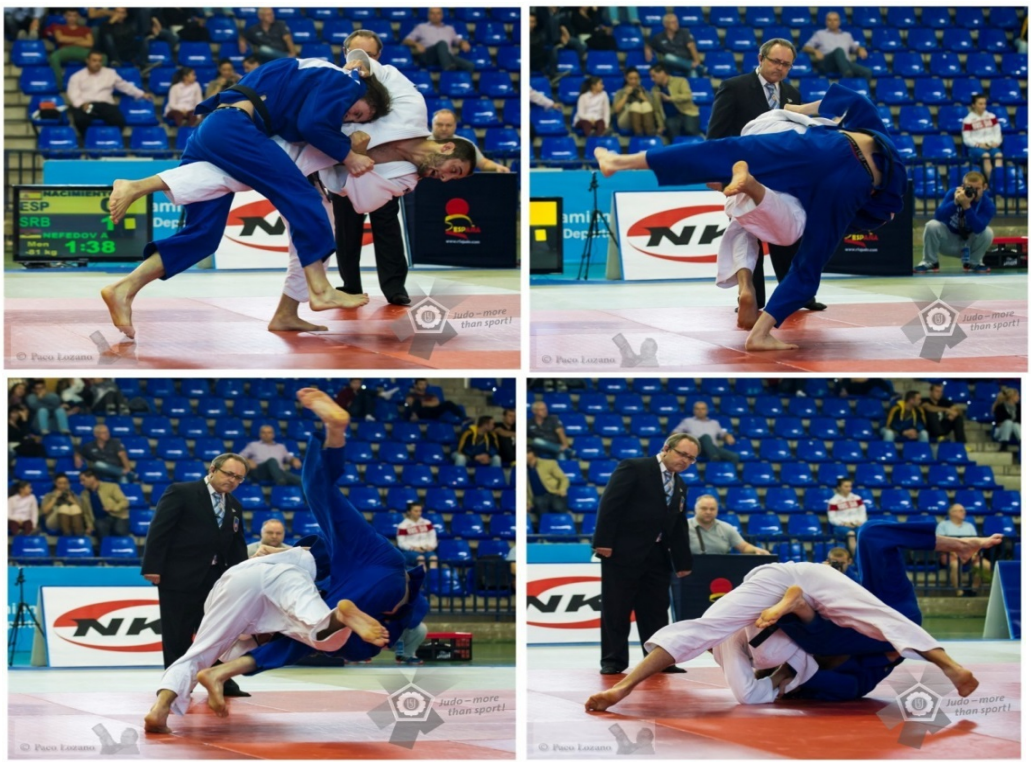

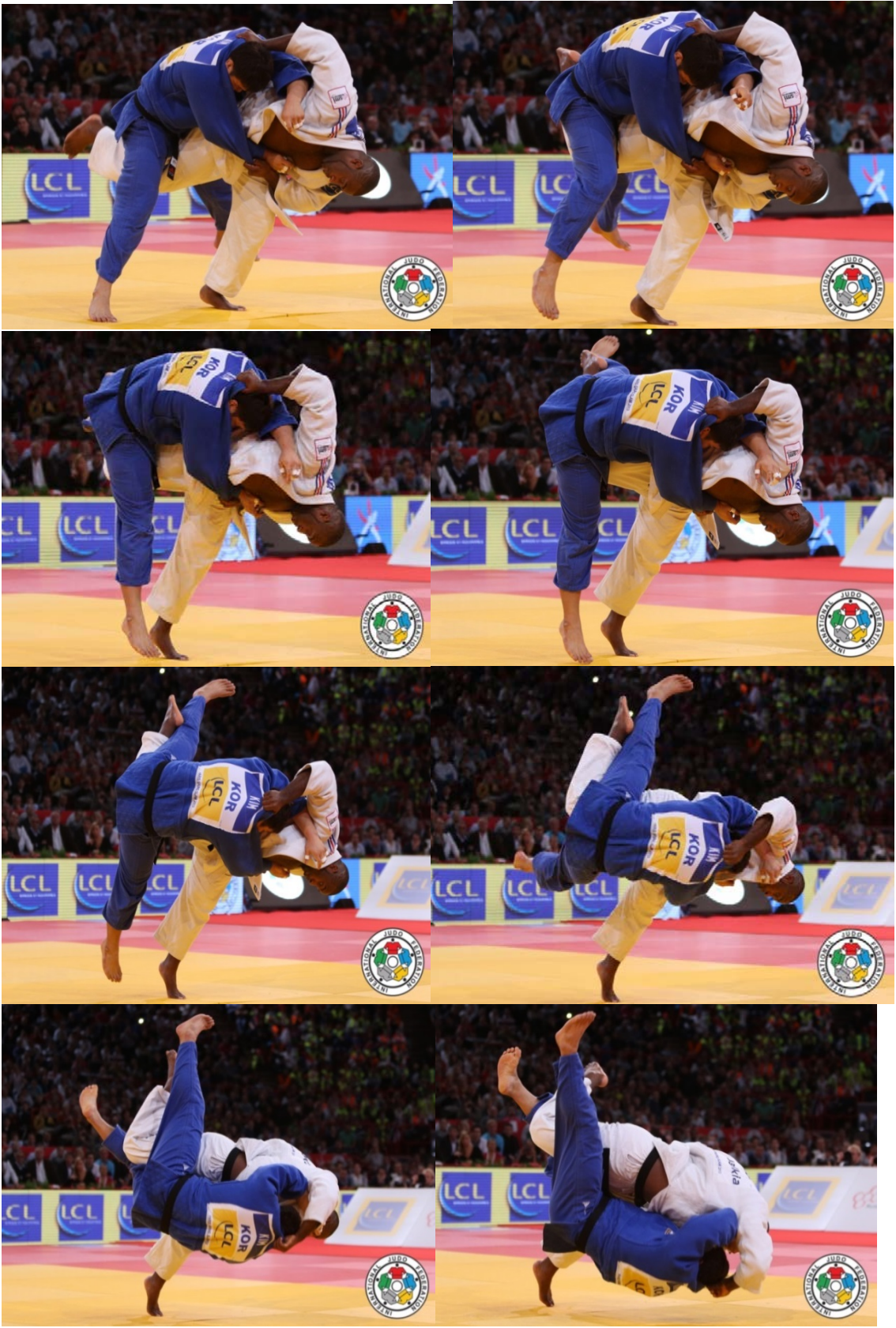

*Fig 35-49 Three different kind of vertical rotational movements in the transverse plan with the axis in the sagittal plane, applied by Tori helping the throwing action in Uchi Mata. Biomechanics helps to infer that the same trick could be applied to O Soto Gari.*

The third way to enhance the Couple tool effectiveness, in competition, is to drive the Uke center of Mass along circular and pseudo-circular paths. Remembering basic mechanics, application of Couple tool insures that the Uke's body moves around his Center of Mass, without translation in space.
Center of Mass movement in space is the trait of Lever application, that biomechanically speaking is the application of a Torque and not a Couple on the Uke's body.
To carry out the third statement means to change the inner mechanics of this class of Judo techniques.
How it is possible to realize this change? Easily in two ways!

1) Mechanics of technique can be changed from Couple to Lever putting back one of forces of Couple. Simultaneous application on the body of equal forces makes a Couple, different time application of the same equal forces on the body makes a Torque. Then translating the Center of Mass along some pseudo-circular paths applying in two times the forces of Couple can enhance effectiveness of some techniques. **Fig 50,55**

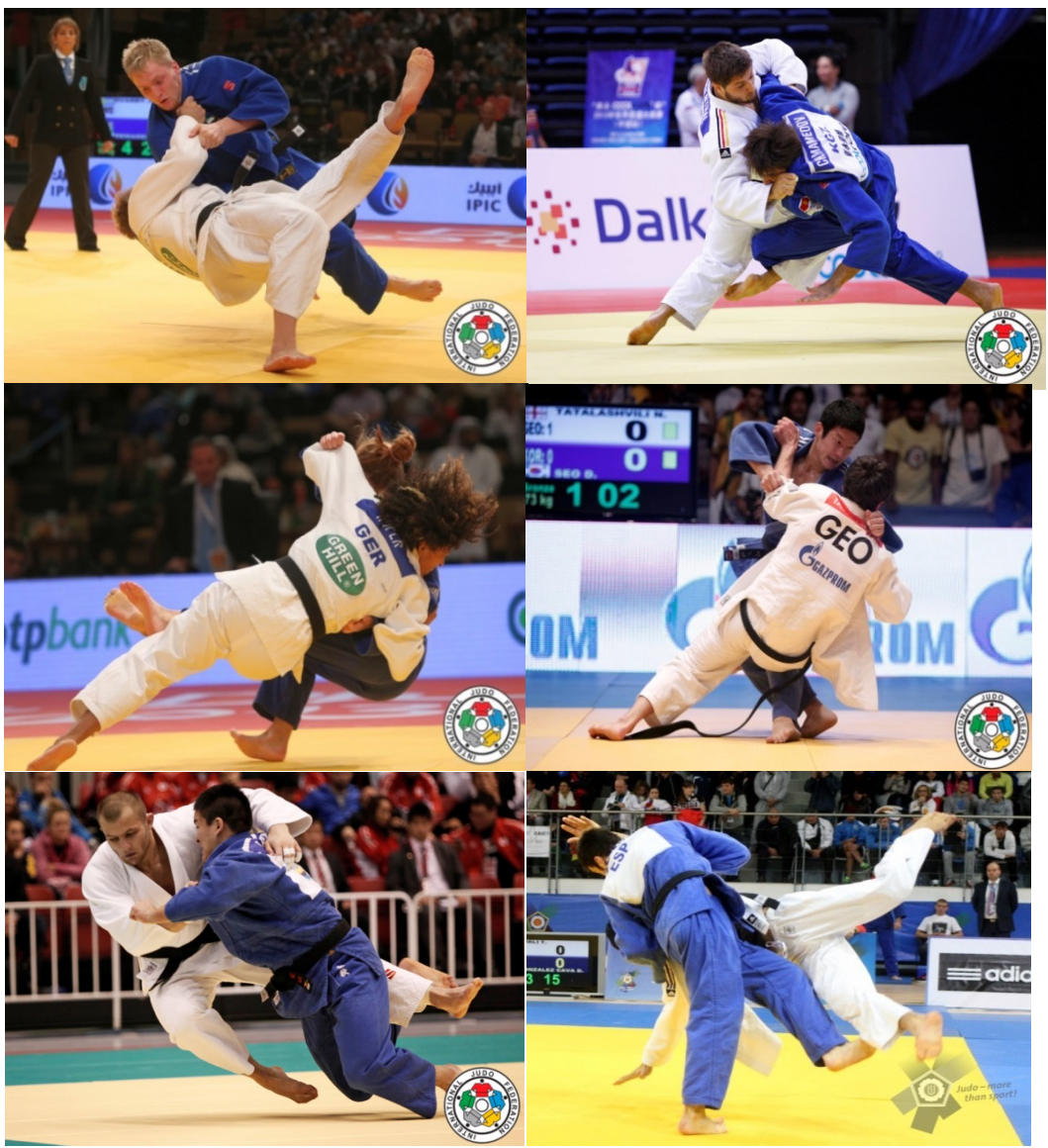

***Fig50,55 Application in two time of Couple forces, drives Uke Center of Mass along a pseudo-circular path in very effective rotational Lever throw born both of De Ashi Arai and of O Uchi Gari.***

In effect, remembering that Couple is the simultaneous application of two parallel forces equal and opposite, if the Athlete applies the two forces in non-parallel way or with a sensible time delay, the mechanics of throws changes from Couple to Lever.
This happens specially with the application of rotational enhancement in competition with vertical rotation in the transverse plane and symmetry axis in the sagittal plane ( rotational application to techniques like Ko Uchi Gari , O Uchi Gari, Okuri Ashi, De Ashi, etc.) , that more often are applied instinctively by athletes in high level competitions.

2) Different dynamical situation that help the changing of one of the Couple forces, in a second time, changing the mechanics of techniques must be considered, sometime, not a very effective enhancement of Couple, but a variation arising from the specific dynamics of the situation.

    Mechanics of technique can be changed from Couple to Lever changing also one of Couple force direction. ***Fig 56,57***

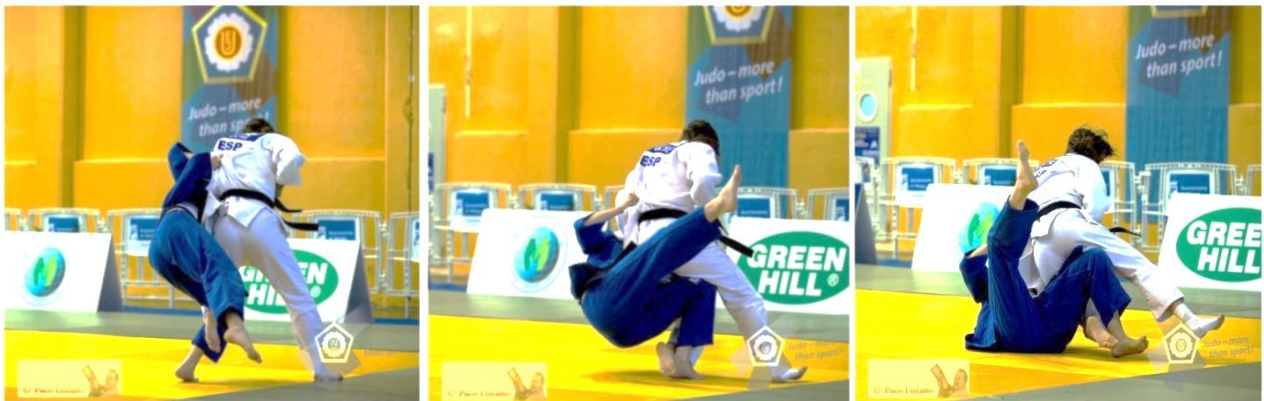

*Fig 56,57 Application of diagonal attack and changing direction of one Couple force- changes from Couple O Soto Gari attack, to lever O Soto Otoshi attack .*

## 7 Rotational variation of Lever techniques

Essentially Couple techniques in competition are based on the "simple" and toneless application of Couple by arms and legs in the human body's three plane of symmetry. Energy
Rotational application in Lever techniques, due to their mechanical essence, is more complex than application in Couple techniques, because in Lever system there are more degree of freedom.
For these kind of techniques the complexity of movement and the mechanics of throws flow also in a higher consumption of energy
A correct biomechanical approach suggests that, because interaction between athletes is a process inside the Couple of Athletes System, it is necessary to break the system and to analyze Tori and Uke as separate bodies.
In such meaning it necessary consider the possible rotation axis connected to the two bodies.
Then to apply a rotational variation of Lever technique it means for Tori the opportunity to rotate in many different directions and symmetry planes to take vantage of the weakened Uke's defensive capabilities.
The play is ruled by the Tori arms that induce different distance between the two bodies.
In such way it is determined not only the rotational mass of the system ( Momentum of Inertia)
$M = m \times r^2$ but also the connected force and the energy.
Essentially lever techniques in competition are based on the following rotation movement that can be changed in inclination bending the body along the 360° in the transverse plan *Fig 58*

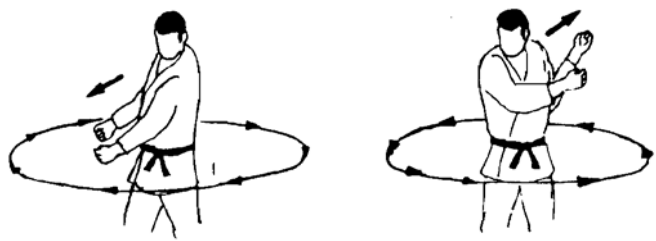

*Fig 58 Basic Rotatory movement that are connected to Lever techniques* [13]

The basic application of rotational movement in Lever techniques group , is always present in classical throws, considering Tori plane of symmetry taking a firm contact point ( fulcrum ) on Uke's body, they are :
A. Vertical Rotation , with the axis in the sagittal plane
B. Transverse Rotation ,with the axis in the frontal plane
C. Antero-posterior Rotation, with the axis in the transverse plane

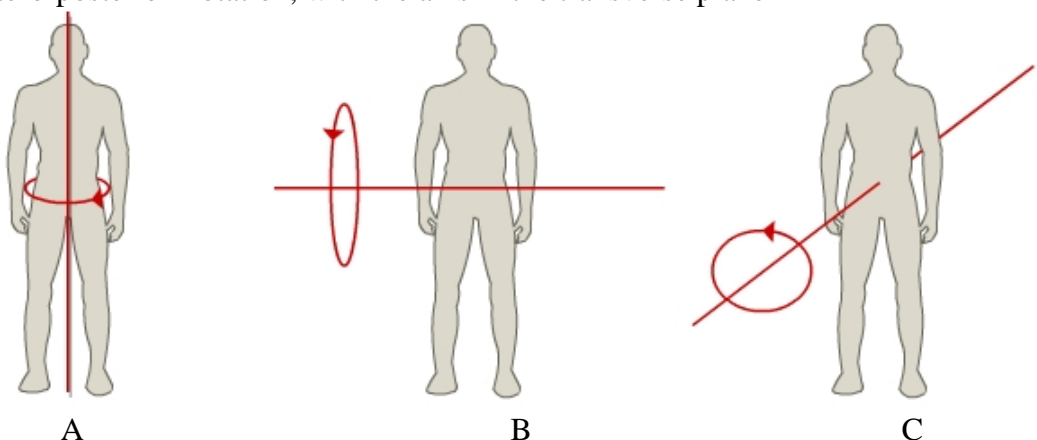

A          B          C
*Fig 59 Possible Rotations and related axis*

Normally the most of classical rotational application flow from the Rotation of Tori around his perpendicular symmetry axe lying in the Sagittal Plane of symmetry.
That is classified as vertical rotation in the previous figure A.
Most of Classical judo Throwing techniques belong to this group, among them is :
O Goshi *Fig 60,61*.
Different axis of rotation is present in well-known classical trick , application of a consecutive Antero-posterior rotation C with the axis in the transverse plane this trick called in Japanese language Makikomi is applied to some throw to enhance the projection effectiveness , see the next group of figures *Fig  62, 65.*

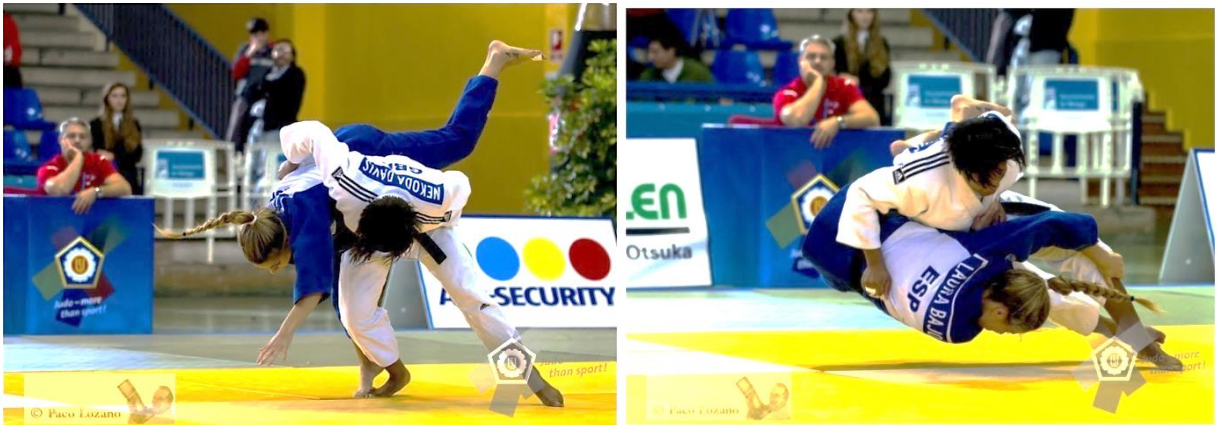

*Fig. 60,61  Vertical Rotation in Classical O Goshi*

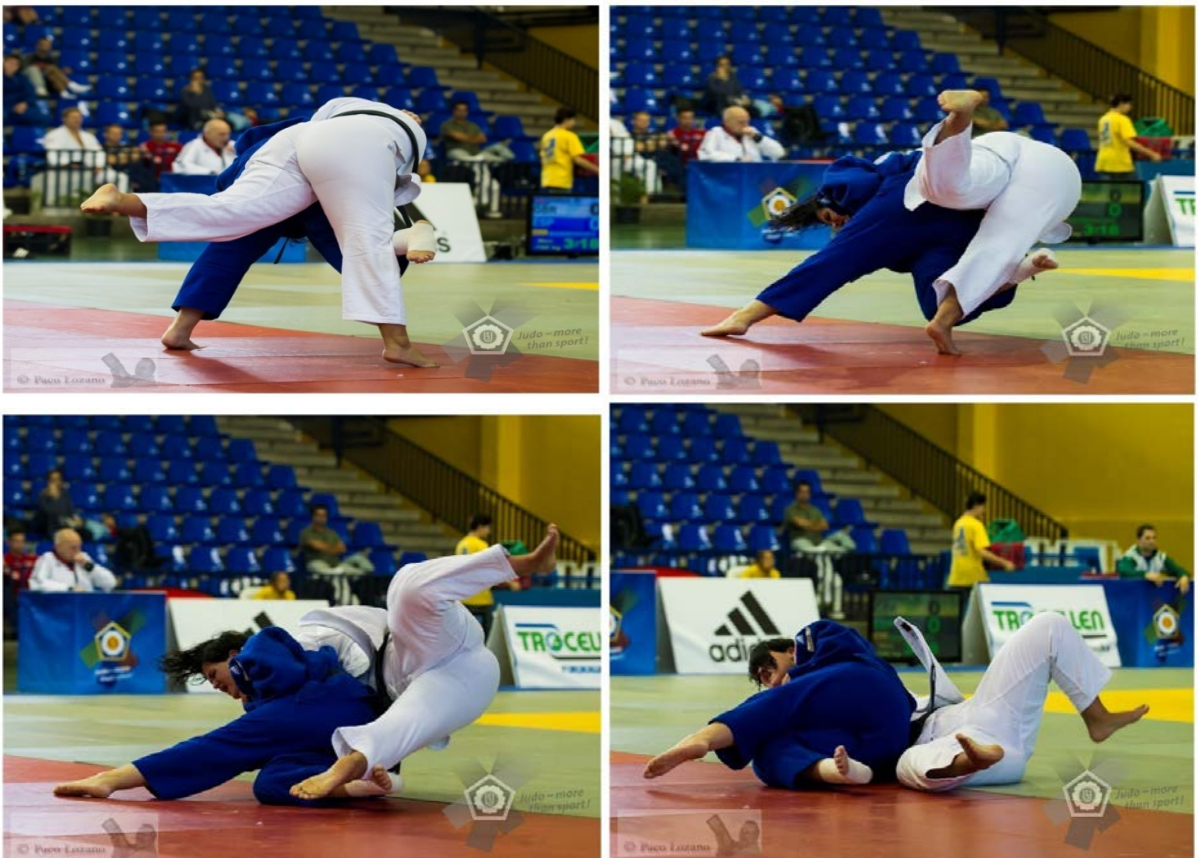

*Fig. 62,65  Antero-posterior rotation in Classical Harai Makikomi*

## 7.1 Enhancement of Lever techniques

From the previous analysis it is clear that the most of Lever judo techniques are inner connected to rotation movements, due to refereeing evaluation regulations, then to enhance effectiveness of Lever techniques it is possible to choose two main way,
1. Hybridization of Lever mechanics introducing a Couple instead of a Torque changing in this way the energetic of technique.
2. Application of rotational dynamics in a totally new way, preserving the lever mechanics

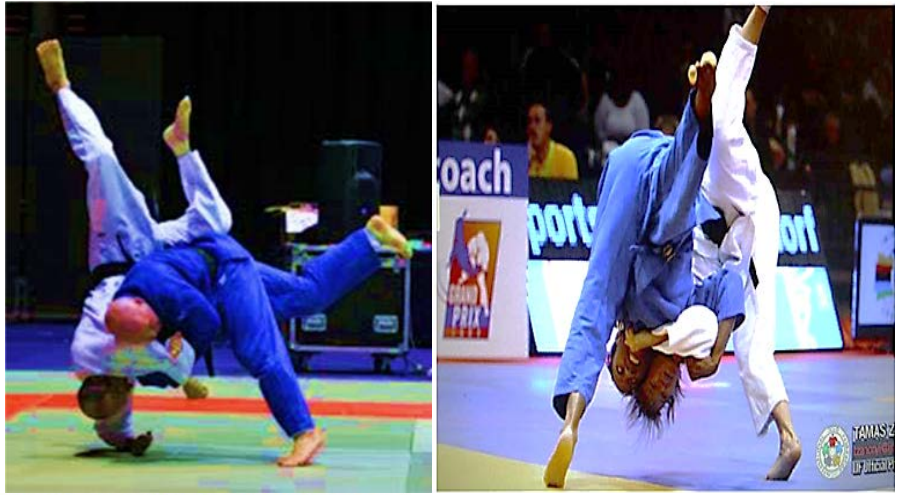

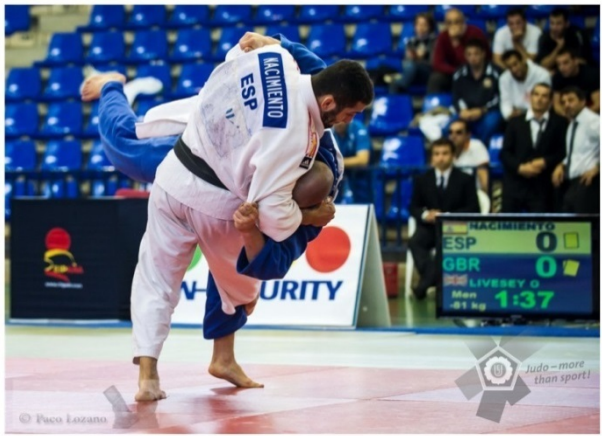
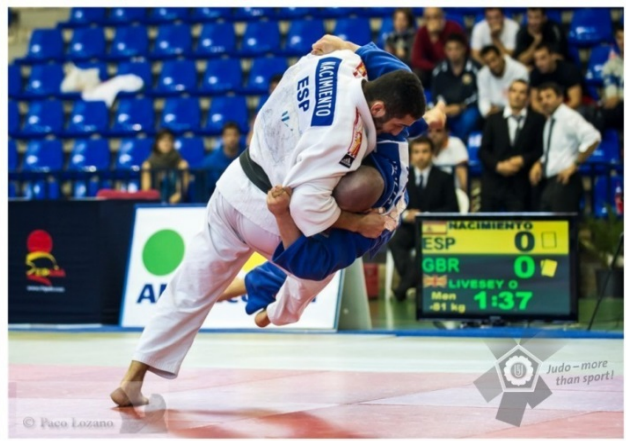

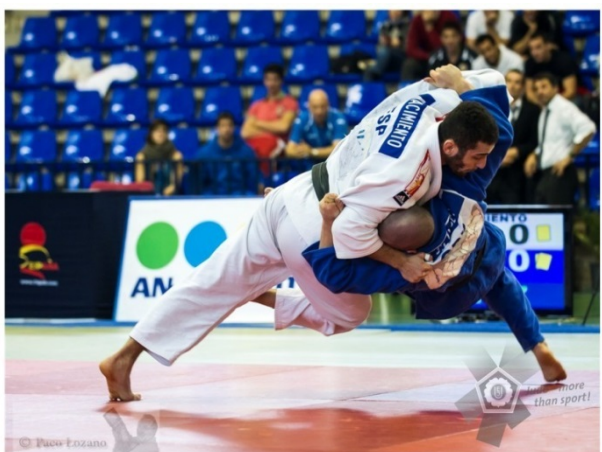
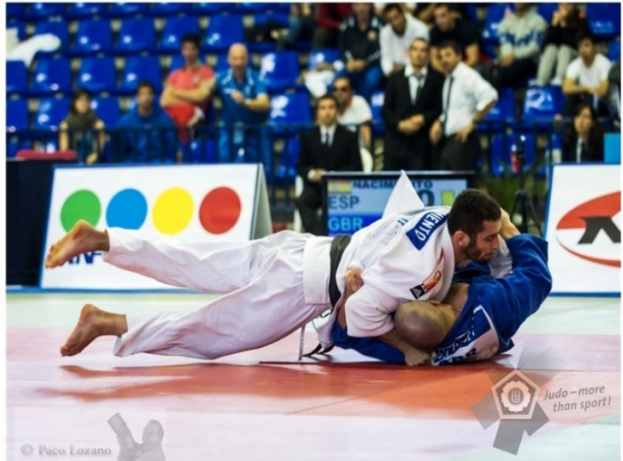

*Fig 66-71 Hybridization with Couple tool of Lever techniques : Morote Seoi, Ippon Seoi and Hiza Guruma, makes the Lever tool more effective and less expensive in energy terms.*

The second method to enhance effectiveness is very interesting and it is grounded on the full-strength rotational application, like the Ueshiba's method for Aikido.
Even if the whole mechanics of the technique does not change at all, the techniques are applied in totally different way, both more effective and less energy wasting.
The first important improvement is that they do not need of unbalance, the most important thing in these new style Lever techniques is to tie tightly with a contact point the adversary's body and to turn swiftly in one of the three plane of reference keeping intact the contact point.
As example the next figure *Fig 72* shows the application of the basic and totally rotatory idea at the classic throws called Morote Seoi Nage, that in this variation could be called as understanding attempt "*Morote Seoi Nage Guruma*", this vertical rotation with the axis in the sagittal plane allows Uke's body fall without unbalance, which is unthinkable in classical application of Lever techniques, and very useful in children teaching and training because doesn't charge the spine with dangerous heavy weight. In the following *fig 73-87* there are shown two application of pure rotation:
The first one applying whatever stopping point during rotation to throws; it is possible to see a pure application of this principle in a limit hip throw at the following hyperlink on you-tube: http://www.youtube.com/watch?v=bZGTAVATPoI .
The second one applying a pure rotation, keeping firmly the contact point, like some Aikido throws.

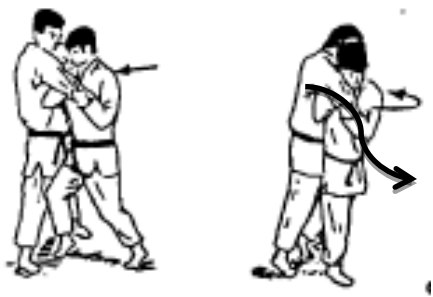

*Fig 72 Morote Seoi Nage (Guruma)  [25]*

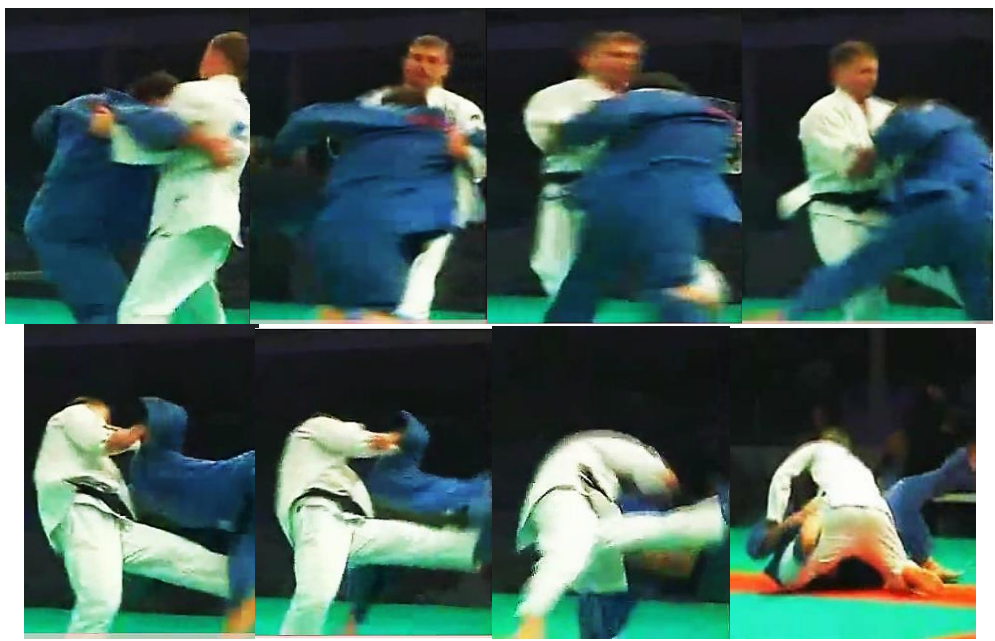

*Fig. 73-80 Fast Vertical Rotation , with the axis in the sagittal plane, utilizing a leg as stopping system:  Chaotic throw - no name.*(it is possible to find a less effective variation of this throw in the Daigo text [26 ]; page. 122)

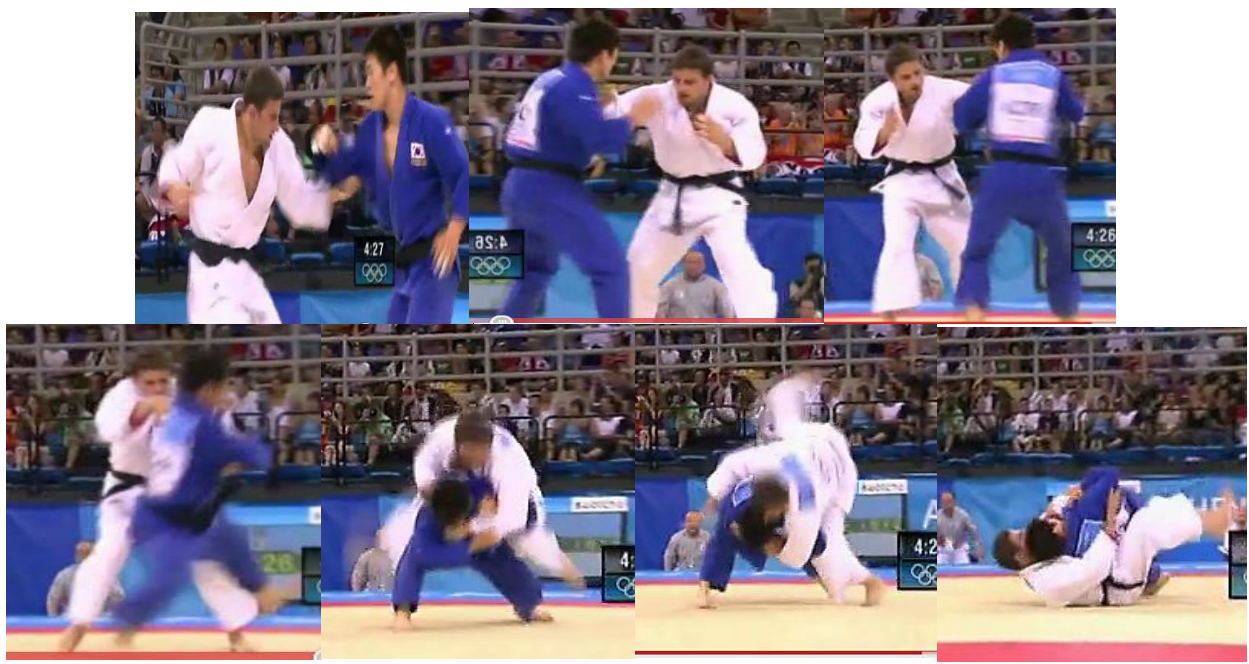

*Fig. 81-87 Fast Antero-posterior Rotation, with the axis in the transverse plane, binding tightly only one arm. Mechanics Lever ( point of application of force left arm, stopping point friction under feet) this technique is structured with the same mechanics of some Lever Throw from Aikido, like : Kata te dori ko kyu nage.*

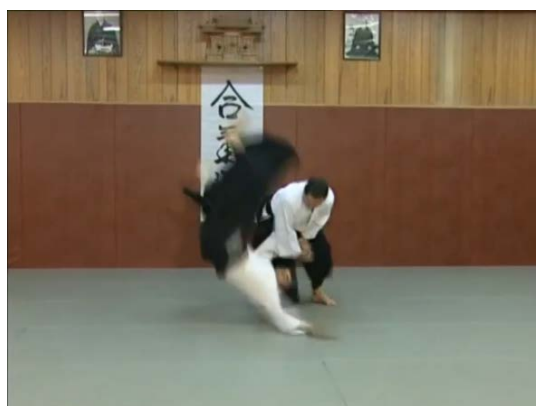

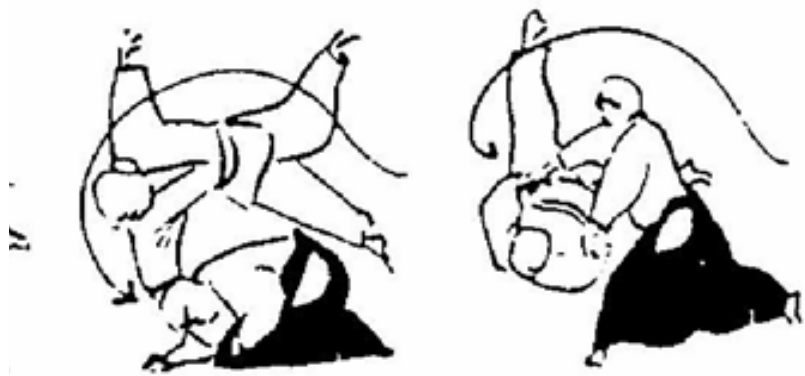

*Fig 88 ,90 Some of the throwing responses form from Kata Te Dori Ko Kyu Nage* [27]

## 8 Physical and Biomechanical framework of the rotational techniques

In classical Newtonian physics the complex rotational application of judo throwing techniques is, for Tori as observer, a clear study of Forces applied in a rotating reference frame.

At first it is important to evaluate the velocity transformation formula from the inertial to the rotating frame, this means how the speed is evaluated by people (public as observer) and Tori as observer during the execution of a rotating throws.

It is essential to find how body's velocity turns into the two frames to evaluate the rule for forces as well. Suppose an athlete (Uke) has position vector *r,* relative to inertial reference frame *O* and position vector *r'* relative to rotating frame *O'*; it follows from the triangle law that :

$$r = D + r' \quad (2)$$

where *D* is the position vector of F' relative to r
the time derivate is:

$$v = V + \left(\frac{dr'}{dt}\right)_O = V + (v' + \omega \wedge r') \quad (3)$$

This is the velocity transformation formula connecting the true and apparent velocities of the tourniong athletes (Uke) as seen by public and by rotating Tori. This formula has a simple interpretation. The true velocity *v* is the sum of:

1. the apparent velocity *v'* in the rotating reference O',
2. the velocity **V+ω ^ r'** that is given to the Uke's body by the motion induced by Tori into the frame *O'*.

An immediate consequence of the velocity transformation formula is the corresponding transformation rule for the acceleration that is the final step to evaluate the force transformation formula .

$$\ddot{r} = A + \alpha \wedge r' + 2\omega \wedge v + \omega \wedge (\omega \wedge r') + a' \quad (4)$$

Now A is the translational acceleration, α the angular acceleration if present in O', and ω the angular velocity of Uke as seen by Tori relative to the inertial point of view ( public).

Then from the previous transformation formula it is possible to evaluate the force transformation:

$$m\ddot{r} = mA + m\alpha \wedge r' + 2m\omega \wedge v' + m\omega \wedge (\omega \wedge r') + ma' \quad (5)$$

In this formula only two fictitious forces are named the third and the fourth, the second one is a rotational acceleration which depends explicitly on the time dependence of the rotation angular velocity. The final equation can be made to resemble the *standard* form of the Newton's Second Law

$$m\frac{d^2r'}{dt^2} = m\frac{\partial^2 r}{\partial t^2} \pm 2m\omega \wedge \frac{\partial r'}{\partial t} \pm m\omega \wedge (\omega \wedge r') \pm m\alpha \wedge r' \pm ma'$$

*the last two fictious terms are present in nonuniform rotation,*
*if we consider them negligible, we have:*

$$m\frac{d^2r}{dt^2} = F_1 + F_{Coriolis} + F_{centrifuga} \quad (6)$$

$$F_{coriolis} = -2m\omega \wedge \frac{dr'}{dt} = -2m\omega \wedge v' \quad (7)$$

$$F_{centrifugal} = -m\omega \wedge (\omega \wedge r') \quad (8)$$

For which into the inertial reference frame every observer ( public) can assess the Newton's Formula applied by Tori to execute the rotational variation of throws, otherwise from the other side into the rotating reference frame the observer connected to the frame ( Tori) when applies a throw will observe a more complicated expression of his force utilized for throwing.
This more complex force for throwing will present, in general, more or two major fictitious contributions, one ordinary called Coriolis contribution, and the other one usually called centrifugal/centripetal contribution. Centripetal and centrifugal contributions are the same force with only the direction difference, but for inertial observer centripetal component is part of the total force  **F = F**$_{Tangential}$ + **F**$_{Centripetal}$, and for rotating observer is a new one , however:

$$F_{centripetal} = m\omega \wedge (\omega \wedge r) \quad \text{inertial observer} \quad (9)$$

$$F_{centrifugal} = -m\omega \wedge (\omega \wedge r) \quad \text{non inertial observer} \quad (10)$$

Then it is possible to write considering only the forces applied on Uke the different vision of the two system of reference:

1. Inertial reference frame for Public: Uke under a rotational attack undergone, is submitted to the following force

$$F = F_{tan\,gential} + F_{Centripetal} = ma \quad (11)$$

2. Non Inertial reference frame for Tori himself, Uke under the same rotational attack is submitted to the following more complex forces

$$F = ma - 2m\omega \wedge v - m\omega \wedge (\omega \wedge r) = ma - (F_{Coriolis} + F_{centrifugal}) \quad (12)$$

The two observations are equal in numerical result, but the physics in the rotational frame is more complex due to the presence of "fictitious forces". It is important to bear in mind that both the Coriolis and Centrifugal forces are at root kinematic effects, resulting from the insistence on using a rotating frame of reference.
In general it is possible to change reference system from the rotating non inertial to the fixed inertial, nevertheless, the transformation between the two frames is usually so complicated that it is easier to work often in the rotating frame and to live with the "fictitious" Coriolis and Centrifugal forces.[28,29,30,31,32,33,34]

## 8.1 Rotating Trajectories

The problem of trajectories of system motion in competition, as already demonstrated, is difficult to analyze.
As found by Scafetta and co, [35] human locomotion is a multi-fractal system (in time and space).
As analyzed in other papers, Judo locomotion is a fractal based point process, and the trajectories made by the Couple of Athletes system CM projection on the mat (or ground tracks) are fractal trajectories created by a fractional Brownian motion [36].
However it is right to analyze a short stretch of trajectories produced by foot in very good approximation by Newtonian physics.
Normally, in real competition, the rotational application of judo throws is restricted only at the first or maximum at the second stride, then the piece of trajectory is in the previous meaning and it is right to use classical tools.
Couple of Athletes system is, considering friction negligible, a free system this entails that conservation of angular system assures us that the motion is planar motion.
Effectively the *moment about the origin* of a force F acting on an Athlete at position r (more or less the length of Tori arms) is, the already discussed Torque that is a vector product

$$T = r \wedge F. \qquad (13)$$

The components of the vector T are the moments about the x-, y- and z-axes,

$$T_x = yF_z - zF_y, \quad T_y = zF_x - xF_z, \quad T_z = xF_y - yF_x. \qquad (14)$$

The direction of the vector T is that of the normal to the plane of r and F, which is the plane of vertical rotation of Couple of Athletes with the symmetry axis into sagittal plane or a plane parallel to the tatami plane.
It may be regarded as defining the axis about which the force of Tori tends to rotate the Uke body.
The magnitude of T is

$$T = rF \sin\theta = bF = armF \qquad (15)$$

where $\theta$ is the angle between r and F, ( Tori arms and line of application of force for throwing) and b is the perpendicular distance from the origin to the line of the force( length of Tori arms).
Correspondingly, we define the vector angular momentum about the origin of a whole body of an Athlete at position r, and moving with momentum p, is

$$J = r \wedge p = r \wedge m\dot{r} = r \wedge mv \qquad (16)$$

Its components, the angular momenta about the x-, y- and z-axes, are

$$J_x = m(y\dot{z} - z\dot{y}), \quad J_y = m(z\dot{x} - x\dot{z}), \quad J_z = m(x\dot{y} - y\dot{x}). \qquad (17)$$

The momentum p is the linear momentum if we have to emphasize the distinction between p and J.
The rate of change of angular momentum, obtained by differentiating (16) is

$$\dot{J} = m\frac{d}{dt}(r \wedge \dot{r}) = m(\dot{r} \wedge \dot{r} + r \wedge \ddot{r}) \qquad (18)$$

Now, the first term in (18) is zero, because it is the vector product of a vector with itself, the second term is simply

$$\dot{J} = r \wedge F \qquad (19)$$

the important result that the rate of change of the angular momentum is equal to the moment of the applied force:

$$\dot{J} = T \qquad (20)$$

This is the rotational analogue of the equation p˙ = F for the rate of change of the linear momentum. The force applied by Tori arms on Uke in a rotational throws application should be considered as an external force applied on Uke, this force is said to be central because it is always directed towards or away from a stable point into the trunk of Tori, called the center of force. (Centrifugal or Centripetal force applied by arms of Tori on Uke Body) If we choose the origin to be this center, this means that F is always parallel to the position vector r. Since the vector product of two parallel vectors is zero, the condition for a force F to be central is that its moment about the center should vanish:
T = r ∧ F = 0.                                                             (21)

From (10) it follows that if the force is central, the angular momentum is a constant:

J = constant.                       (22)

This is the law of conservation of angular momentum.

It really contains three statements:

1. the direction of J is constant,
2. its magnitude is constant
3. the Torque lies into a plane parallel to mat plane

It is interesting now to apply for practical method the ideal elongation of the short stretch till to complete trajectory that is closed and could be evaluated as two body system better in the approximation of the Bertrand theorem.
This theorem demonstrates that not only the Newtonian gravitation force F= k/ $r^2$ can generate closed orbits but also the Hooke force F= - k r can do the same.
In our approximation it is better to choose one Hooke like force, because Tori's arms are better approximated by a spring than by a gravitational attraction.
Deriving by Lagrangian of the system, the trajectory equation is of the form:

$$\frac{l^2}{mr^2}\left[\frac{d^2}{d\theta^2}\left(\frac{1}{r}\right) + \left(\frac{1}{r}\right)\right] = f\left(\frac{1}{r}\right) \qquad (23)$$

The condition for a circular orbit of radius $r_0 = \cos t$
In addition, of course, the energy must satisfy the condition E<0 . If the energy is slightly above that needed for circularity, and the potential is such that the motion is stable, then r will remain bounded and vary only slightly from $r_0 = \cos t$ and f(1/r) can be expressed in terms of the first term in a Taylor series expansion:

$$f\left(\frac{1}{r_0}\right) = \frac{1}{r_0} + \left(\frac{1}{r} - \frac{1}{r_0}\right)\frac{d}{dr_0}\left[f\left(\frac{1}{r}\right)\right] + O^2\left(\frac{1}{r}\right) \qquad (24)$$

The researched solution of the general force is: $f(r) = -\dfrac{k}{r^{3-\alpha^2}}$ and with some mathematical evaluation that it is possible to find in many books of classical mechanics[37,38,39,40] it is possible to demonstrate that the closed orbits can arise from only two solution the Newtonian gravitation force $F = k/r^2$ and the Hooke force $F = -kr$, that is the solution researched.

Some trajectories can be shown by a physical analogue: a spring mass system on a rotating table
In the next figure *Fig 91* there is an example of trajectories using Wolfram "Mathematica" software, physical exercise prepared by Nasser Abbassi 2013.

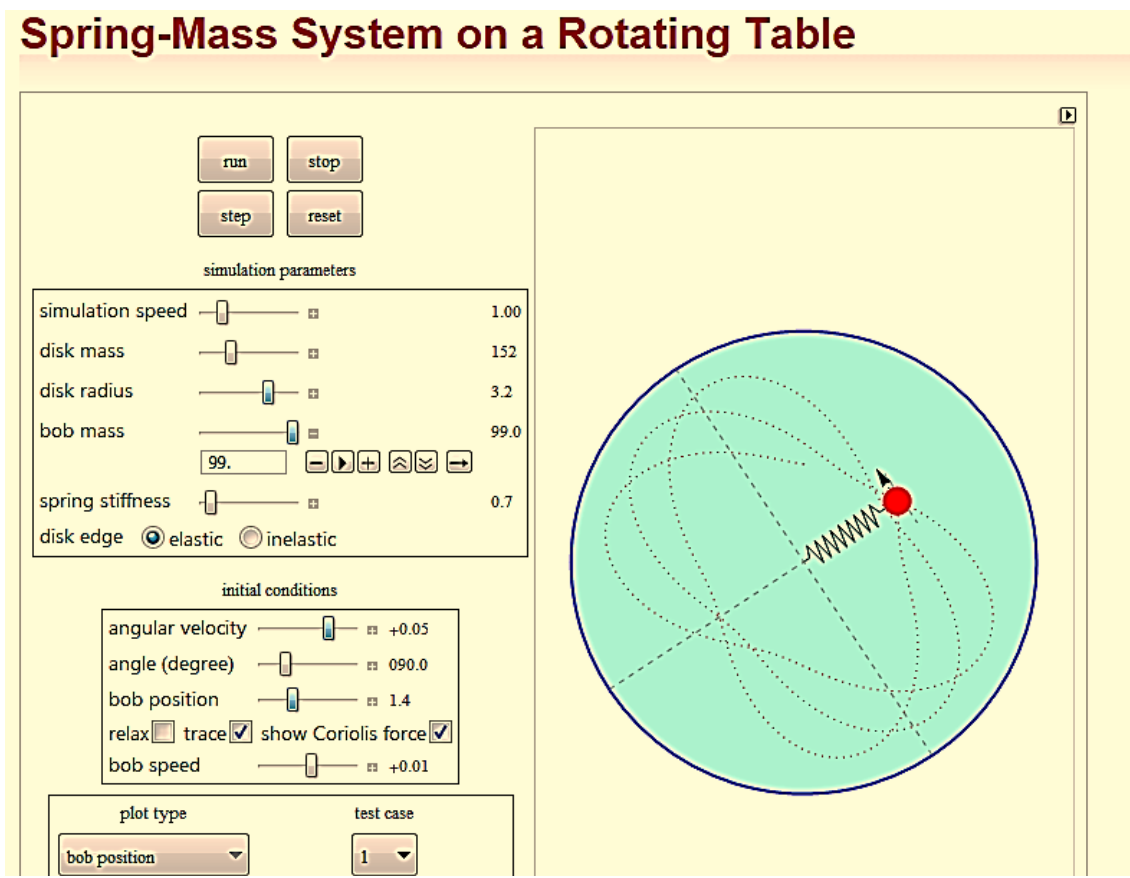

*Fig.91 Rotatory trajectories modelled by a spring –mass on a rotating table.*

http://demonstrations.wolfram.com/author.html?author=Nasser%20M.%20Abbasi [41]

In term of trajectories to enhance the throws effectiveness, some of **General Action Invariants** are almost right lines with specific direction, normally as we see before, the best right inclination is, in direction of both adversaries' sides, because human body structure is less skilled to resist in such direction. For example into the group of couple of forces, this means couple applied in the frontal plane for an antero-posterior rotation with the axis in the transverse plane : like Okuri Ashi (harai – barai).

For the other classes based on rotations, some interesting remarks come from the Poinsot geometrical description of a free forces motion of a body, in such case in fact the motion is like a rolling of the body inertial ellipsoid (without slipping) on a specific plane, remembering that the curve traced out by the point of contact on the inertial ellipsoid is called polhode, while the curve on the plane is called herpolhode.

In a case like judo player, the body is cylindrical symmetrical and the inertial ellipsoid becomes an ellipsoid of revolution, then the polhode is a circle around the Athlete symmetry axis, and the herpolhode on the invariant plane (Tatami) is likewise a circle. In the next *fig 92* the athlete is modeled by a parallelepiped shape.

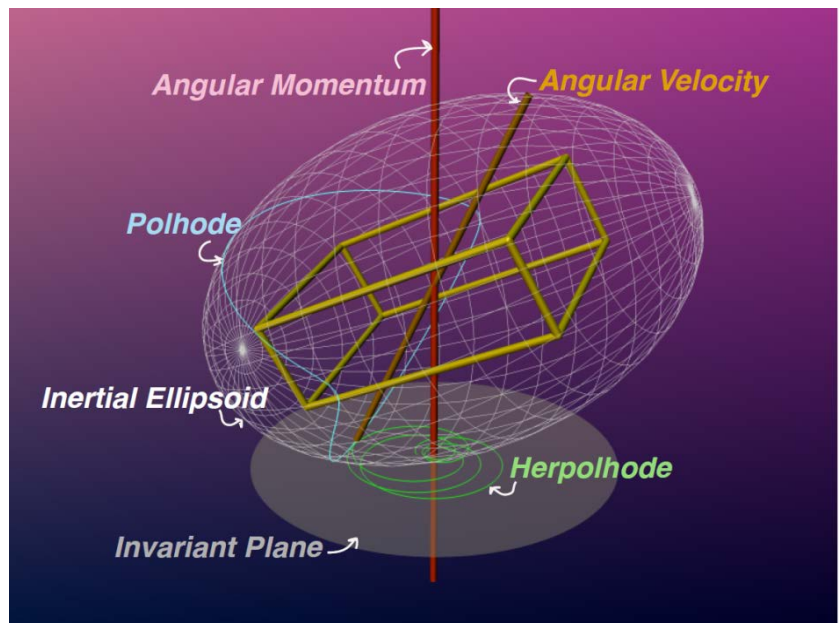

*Fig.92 Poinsot geometrical description of athlete ( parallelepiped) motion during interaction on the tatami* [42]

These results in the judo player reference system are real only in the case of free force motion, (it is true considering the couple motion as a whole) but in our case interaction into the couple, as already remembered, there are the push/pull forces and the friction forces acting on.

The problem is almost not analytically solvable for his complexity. However the gross indication (likewise circular trajectories) available for our analysis by Poinsot description is quite acceptable as indication for real situations. The judo movements have very complex 3D trajectories, and then it is very difficult to calculate the forces time evolution in space.

The resultant mean push/pull force during shifting motion was already evaluated in the first Sacripanti model [43] as:

$$\Delta F(t-t') = \Delta v/m \ \Sigma(\pm 1)\delta(\ t-t') \qquad (25)$$

The previous is a generalized force in a Langevin type Equation. However there are more cashed complexities, like geometrical adaptation of structure of muscles involved, that change the amount of contraction forces and so on, which make the problem not only beyond words but also analytically insoluble.

*9 Conclusion*

In this paper it is faced the problem of the enhancement of throws in judo competition, normally as the pictures show this problem is already solved intuitively by champions during high dynamic situations that arise during competition.
This also because the practical mechanical solution in term of trajectories, movements and forces directions follow the rules found by Newton in his classical mechanics.
The biomechanical analysis performed try to find the basic paradigm of the problem, to improve the technical training focalized on the effectiveness of throws in competition.
The results gathered are summarized in easy and simple steps, as follow.

Lever Techniques can enhance their effectiveness in three ways :

4. The rotational movements, strictly connected to the Lever techniques mechanics achieving victory (Ippon) in competition, can be extended to the unbalance phase (Kuzushi)
5. The rotational movements can be applied in a totally new way putting away even the unbalance that is basic in the Lever techniques.
6. The Lever tool can be hybridized with the application of a Couple to lower the energy consumption and to overcome some strong defensive resistance.

Couple Techniques can enhance their effectiveness in three ways:

4. The Couple tool that, in principle, doesn't need unbalance allowing uke's body rotate around his center of mass, it is enhanced utilizing the Uke's body smaller resistance directions (normally summarized in Diagonal attacks).
5. The vertical rotational movements in the transverse plan with the axis in the sagittal plane can be added to the Couple application with Transverse Rotation, and axis in the frontal plane to overcome some defensive resistance, mainly in the trunk and leg group of Couple techniques ( like Uchi Mata or O Soto Gari).
6. The rotational movements can enhance the throwing action changing the inner mechanics of Couple into Lever applying a Torque, with the direction change of one force or the time delay of his application.

The mathematics explication of the rotation in the Classical Newtonian Mechanics shows both the two point of view the inertial reference and rotating reference system.
As well known, both results are equal in numerical term but the physics forces are different in the rotating frame there are in general four forces added but the most well-known are two the Centrifugal and Coriolis forces.